\documentclass{emulateapj}

\usepackage{lscape}

\slugcomment{Accepted to AJ, 2011 Sep 11}

\shorttitle{Millimeter emissions of nearby early-type galaxies and LLAGNs}
\shortauthors{Doi et al.}

\begin{document}

\title{Millimeter Radio Continuum Emissions as the Activity of Super Massive Black Holes in Nearby Early-Type Galaxies and Low-Luminosity Active Galactic Nuclei}

\author{Akihiro~Doi\altaffilmark{1,2}, Kouichiro~Nakanishi\altaffilmark{3}, Hiroshi~Nagai\altaffilmark{3}, Kotaro~Kohno\altaffilmark{4}, and Seiji~Kameno\altaffilmark{5}}

\altaffiltext{1}{The Institute of Space and Astronautical Science, Japan Aerospace Exploration Agency, 3-1-1 Yoshinodai, Chuou-ku, Sagamihara, Kanagawa 252-5210, Japan}\email{akihiro.doi@vsop.isas.jaxa.jp}
\altaffiltext{2}{Department of Space and Astronautical Science, The Graduate University for Advanced Studies,\\ 3-1-1 Yoshinodai, Chuou-ku, Sagamihara, Kanagawa 252-5210, Japan}
\altaffiltext{3}{National Astronomical Observatory of Japan, 2-21-1 Osawa, Mitaka, Tokyo 181-8588, Japan}
\altaffiltext{4}{Institute of Astronomy, The University of Tokyo, 2-21-1 Osawa, Mitaka, Tokyo 181-0015, Japan}
\altaffiltext{5}{Department of Physics, Faculty of Science, Kagoshima University, 1-21-35 Korimoto, Kagoshima, Kagoshima 890-0065, Japan}

\begin{abstract}
We conducted millimeter continuum observations for samples of nearby early-type galaxies (21~sources) and nearby low-luminosity active galactic nuclei~(LLAGN; 16~sources) at 100~GHz ($\lambda3$mm) using the Nobeyama Millimeter Array~(NMA).  In addition, we performed quasi-simultaneous observations at 150~GHz ($\lambda2$mm) and 100~GHz for five LLAGNs.  Compact nuclear emissions showing flat or inverted spectra at centimeter-to-millimeter wavelengths were found in many LLAGNs and several early-type galaxies.  Moreover, significant flux variability was detected in three LLAGNs.  These radio properties are similar to Sgr~A*.  The observed radio luminosities are consistent with the fundamental plane of black hole activity that has suggested on the basis of samples with black hole masses ranging from 10 to 10$^{10}\ M_\sun$.  This implies nuclear jets powered by radiatively inefficient accretion flows onto black holes.  
\end{abstract}

\keywords{galaxies: active --- galaxies: jets --- galaxies: elliptical and lenticular, cD --- radio continuum: galaxies --- techniques: interferometric}

\section{INTRODUCTION}\label{section:introduction}
Active galactic nuclei~(AGNs) are believed to be powered by gravitational energy released from accreting matter onto supermassive black holes~(SMBHs) at the center of galaxies.  Our galaxy also hosts an SMBH as known as Sgr A* in the galactic center.  Sgr~A* could be detected by an extragalactic observer as a non-AGN, although it is an accretion-powered radio source on an SMBH \citep{Gillessen_etal.2009}.  The spectral energy distributions~(SEDs) of Sgr~A* are explained by accretion models such as advection-dominated accretion flow \citep[ADAF;][]{Narayan_etal.1995,Manmoto_etal.1997} or radiatively inefficient accretion flow \citep[RIAF][]{Yuan_etal.2003}.  The radio spectrum of Sgr~A* is inverted ($\alpha>0$ and $S_\nu \propto \nu^\alpha$, where $\alpha$ is the spectral index, $S_\nu$ is the flux density, and $\nu$ is the frequency).  The millimeter emission is much stronger than the centimeter one \citep{Falcke_etal.1998,An_etal.2005} and is characterized by rapid and large-amplitude flux variability \citep{Miyazaki_etal.2004,Yusef-Zadeh_etal.2011}.  Similar radio properties have been established for M~81 \citep{Reuter&Lesch1996,Sakamoto_etal.2001,Markoff_etal.2008,Schodel_etal.2007}, which is a nearby low-luminosity AGN~(LLAGN) with radio luminosity of more than four~orders of magnitude higher than that of Sgr~A*.  Inverted spectra at centimeter-to-millimeter regimes have been observed in several other nearby LLAGNs \citep{Nagar_etal.2001,Anderson_etal.2004,Doi_etal.2005b}.  These LLAGNs are also powered by sub-Eddington SMBHs.  Thus, Sgr~A* may be one of the commonplace SMBHs in the local universe.  In the centers of elliptical galaxies and the bulges of spiral galaxies in the local universe, there are many starved SMBHs as AGN relics, which have grown by mass accretion during the past AGN phases \citep[e.g.,][]{Marconi_etal.2004}.  Sensitive radio observations established the prevalence of weak activity in nearby early-type galaxies, and that most bright early-type galaxies have low-luminosity nonthermal radio sources at centimeter regimes \citep[e.g.,][]{Sadler_etal.1989,Brown_etal.2011}.  About 70\% of radio-emitting early-type galaxies have compact radio cores, which even at low radio luminosities usually show flat or inverted spectra at centimeter regimes like radio galaxies' cores \citep{Slee_etal.1994}.  However, the millimeter radio properties of these starved SMBHs have been less understood because their weak activities could not be detected with the sensitivities of conventional millimeter radio telescopes.    

The weak activities of these starved SMBHs may be contaminated by the steep spectra ($\alpha<0$) of nonthermal emissions related to star formation at centimeter wavelengths.  Millimeter observations are promising to observe the activities of LLAGNs, as well as Sgr~A*.  Because dust emissions associated with host galaxies could be contaminated at millimeter/submillimeter regimes, observations using millimeter interferometry with the smallest possible beam sizes are crucial.  The detection of variability also helps eliminate the possibility of the contribution of dust emission.  

The present paper reports millimeter continuum observations of the samples of nearby LLAGNs and nearby early-type galaxies using the Nobeyama Millimeter Array~(NMA).  The data will contribute to understanding the accretion phenomena on SMBHs in nearby galaxies and in planning observations with the Atacama Large Millimeter/Submillimeter Array, which is supposed to detect millimeter emissions from much larger number of galaxies at a much higher signal-to-noise ratios.  

The paper is structured as follows.  The samples are drawn in \S~\ref{section:sample}.  In \S~\ref{section:observation}, we describe our observations and data reduction procedures.  In \S~\ref{section:result}, we present the results.  In \S~\ref{section:discussion}, we discuss the physical origin of the detected radio emissions.  Finally, the paper is summarized in \S~\ref{section:summary}.

\section{SAMPLES}\label{section:sample}
We prepared two samples for the NMA observations; one is the sample of nearby LLAGNs for investigating variability, and the other is the sample of nearby early-type galaxies for exploring millimeter emissions.

The nearby LLAGN sample consists of 16~sources (Table~\ref{table1}), selected from 20~sources in the sample of our previous study using NMA \citep{Doi_etal.2005b}; all the 20~LLAGNs have been detected using Very Long Baseline Interferometries (VLBIs) at any bands of 1.4--43~GHz ($\lambda\lambda21$--0.7cm), which guarantees the presence of compact radio components at galactic centers.  The previous study reported flux density measurements at $\sim$100~GHz~($\lambda3$mm) using data integrated over multiple epochs in order to improve sensitivities for the sources.  In this paper, we report the flux densities of each epoch.  NGC~4168, NGC~4203, NGC~4472, and NGC~4552 were excluded because multi-epoch observations had not been performed for these sources.  The well-known source NGC~4486~(M~87) was also not included because the previous study had used its flux density reported in the literature.  We included NGC~6500 in the LLAGN sample.  Although the source was VLBI detected, its luminosity might be slightly higher than the operational definition of LLAGNs \citep[$L_\mathrm{H\alpha}<10^{40}$~erg~s$^{-1}$;][]{Ho_etal.1997a}; thus, the source was not included in the sample of the previous study.  We report the results for this source in this paper.  We also prepared a subsample of five~sources (NGC~3031, NGC~4278, NGC~4374, NGC~4579, and NGC~6500), which are the strongest radio sources in the LLAGN sample, to perform NMA observations quasi-simultaneously at $\sim$150~GHz~($\lambda2$mm) and $\sim$100~GHz~($\lambda3$mm).      

The nearby early-type galaxy sample consists of 21~sources (Table~\ref{table2}) selected from the early-type galaxies with dusty nuclear disks or filaments detected using the {\it Hubble Space Telescope} \citep[{\it HST};][]{Tran_etal.2001}.  \citet{Tran_etal.2001} conducted observations for two major samples.  The main sample consisted of E or S0 galaxies (67~galaxies in total) from the Lyon/Meudon Extragalactic Database~(LEDA), selected to be nearby with $v<3400$~km~s$^{-1}$, lying at Galactic latitude $>20\degr$, and observed by {\it HST} using WFPC2 in the snapshot mode with the F702W filter~(``snapshot sample'').  The other sample consisted of E or S0 galaxies (40~galaxies in total) with selection criteria similar to those of the snapshot sample except that the galaxies were drawn from the {\it HST} archive and with the additional criterion that a 100~$\mu$m {\it IRAS} detection existed at $\ga 3 \sigma$~(``{\it IRAS} sample'').  Dust features were detected in 29 and 30~galaxies for the snapshot and {\it IRAS} samples, respectively.  The dust-detected galaxies were categorized into four dust levels \citep{Rest_etal.2001}: (1)~filamentary low, (2)~filamentary medium, (3)~filamentary high, and (4)~dusty disk.  The level of filamentary dust was described in a purely qualitative way, with the simple scale 1, 2, or 3 being assigned to denote the least to most dusty galaxies through visual inspection only.  Class 1 represented small traces of dust that do not greatly affect the isophotal shape of the galaxy; classes 2 and 3 denoted large amounts of dust that prevent a meaningful analysis of the isophotes and luminosity profiles.  We selected 21~sources out of the 29+30 dust-detected galaxies with the following selection criteria: exclusion of dust level~(1), detection of optical nuclear line emissions, declination of $>-10\degr$, and the exclusion of radio galaxies with 3C-like structures.  As a result, NGC~4168 and NGC~4552, which had been excluded from the LLAGN sample because of no multi-epoch data, were restored in the nearby early-type galaxy sample.  In contrast, NGC~3226 was excluded from the early-type galaxy sample because this source had already been included in the LLAGN sample.  NGC~4278 and NGC~4374 with significant dust features were excluded from the early-type galaxy sample because they are radio galaxies; however, they had already been included in the LLAGN sample.  After our observations, the VLBI detections of several galaxies in the early-type galaxy sample have been reported: NGC~4589, NGC~5322 \citep{Nagar_etal.2005}; NGC~5077 \citep{Petrov_etal.2006}; and NGC~5846 \citep{Filho_etal.2004}.

\section{OBSERVATIONS AND DATA REDUCTION}\label{section:observation}
To study variability, we reanalyzed the NMA data for LLAGNs obtained in our previous study \citep{Doi_etal.2005b} and divided them into each epoch.  Almost all sources, except NGC~4258, were observed using the same instrumental parameters.  We used the D configuration, which is the most compact array configuration resulting in half-power beam widths (HPBWs) of synthesized beams of $\sim 7 \arcsec$.  The observations spanned more than 20 observation days between 2002 November~28 and 2003 May~25 (one exceptional observation for NGC~2787 was conducted on 2004 April~25).  Visibility data were obtained with a double-sided-band receiving system at center frequencies of 89.725 and 101.725 GHz, which were Doppler-tracked.  We used the Ultra Wide-Band Correlator \citep[UWBC;][]{Okumura_etal.2000}, which can process a bandwidth of 1~GHz per sideband, i.e., 2~GHz in total.  Such frequency bands can avoid possible contamination of all sources in the sample from several significant line emissions: $^{13}$CO($J=1-0$), C$^{18}$O($J=1-0$), HCN($J=1-0$), HCO$^{+}$($J=1-0$), and SiO($J=2-1$).  A system noise temperature, $T_\mathrm{sys}$, was typically about 150~K in the double-sided bands.  For observations of each target, we scanned a reference calibrator near the target every 20 or 25~minutes for amplitude and phase calibration.  Bandpass calibrators of very bright quasars were scanned once a day.  Flux scales of the gain calibrator were derived on each observation day with uncertainty up to 15\% by relative comparisons to flux calibrators such as Uranus, Neptune, or Mars with a known flux density.  The flux calibrators were quasi-simultaneously observed at almost the same elevations in order to avoid residual errors in calibration of differential atmospheric attenuation.  For NGC~4258, data were obtained with all configurations~(AB, C, and D); HPBWs of the synthesized beams were typically $\sim 3 \arcsec$, $\sim 5 \arcsec$, and $\sim 7 \arcsec$ in the AB-, C-, and D-array configurations, respectively.  Observations were originally conducted for HCN($J=1-0$) and HCO$^{+}$($J=1-0$) emission lines, except for those between 2005 May~13 and 15, which were only for continuum emission.  In the former data, we used only the line-free band of the upper sideband at a center frequency of 100.777~GHz.  In the latter data (between 2005 May~13 and 15), we used both the sidebands of 89.729 and 101.729~GHz.

\defcitealias{Tully1988}{1}
\defcitealias{Tonry_etal.2001}{2}
\defcitealias{Freedman_etal.1994}{3}
\defcitealias{Herrnstein_etal.1999}{4}
\defcitealias{Solanes_etal.2002}{5}
\defcitealias{Ho_etal.1997a}{6}
\defcitealias{Roberts_etal.1991}{7}
\defcitealias{Bettoni_Buson1987}{8}
\defcitealias{Goudfrooij_etal.1994}{9}
\defcitealias{Kollatschny_Fricke1989}{10}
\defcitealias{Gonzalez1993}{11}
\defcitealias{Huchra_Burg1992}{12}
\defcitealias{Doi_etal.2005a}{13}
\defcitealias{Kormendy_Gebhardt2001}{14}
\defcitealias{Ho_etal.1997b}{15}
\defcitealias{Sarzi_etal.2001}{16}
\defcitealias{Heckman_etal.1980}{17}
\defcitealias{Devereux_etal.2003}{18}
\defcitealias{McElroy1995}{19}
\defcitealias{Eracleous_etal.2010}{20}
\defcitealias{Ho2007}{21}
\defcitealias{Tremaine+2002}{22}
\defcitealias{Pellegrini2010}{23}
\defcitealias{Bower_etal.1998}{24}
\defcitealias{Doi_etal.2005b}{25}
\defcitealias{Bender_etal.1994}{26}
\defcitealias{Bernardi_etal.2002}{27}
\defcitealias{Gultekin_etal.2009}{28}
\defcitealias{Gebhardt_etal.2003}{29}
\defcitealias{Wegner_etal.2003}{30}
\defcitealias{Kuntschner_etal.2001}{31}
\defcitealias{Terashima_Wilson2003}{32}
\defcitealias{Ho_etal.2001}{33}
\defcitealias{Gonzalez-Martin_etal.2006}{34}
\defcitealias{Satyapal_etal.2005}{35}
\defcitealias{Reynolds_etal.2009}{36}
\defcitealias{Halderson_etal.2001}{37}
\defcitealias{Eracleous_etal.2002}{38}
\defcitealias{Komossa_etal.1999}{39}
\defcitealias{Liu_Bregman2005}{40}
\defcitealias{Iyomoto_etal.1998}{41}
\defcitealias{OSullivan_etal.2001}{42}
\defcitealias{Desroches_Ho2009}{43}
\defcitealias{Panessa_etal.2006}{44}
\defcitealias{Soria_etal.2006}{45}
\defcitealias{Ghosh_etal.2007}{46}
\defcitealias{Pellegrini2005}{47}

	\renewcommand{\arraystretch}{0.7}

\begin{table*}
\caption{Nearby LLAGN Sample\label{table1}}
\begin{center}
\begin{tabular}{lrccccccccc} \tableline\tableline
\multicolumn{1}{c}{Name} & \multicolumn{1}{c}{$D$} & Ref. & Dust & Line & Ref. & Morph. & $\log{M_\mathrm{BH}}$ & Ref. & $L_\mathrm{X}$ & Ref. \\
\multicolumn{1}{c}{} & \multicolumn{1}{c}{(Mpc)} &  &  & Emission &  &  & ($M_\mathrm{\sun}$) &  & (erg s$^{-1}$) &  \\
\multicolumn{1}{c}{(1)} & \multicolumn{1}{c}{(2)} & (3) & (4) & (5) & (6) & (7) & (8) & (9) & (10) & (11) \\\tableline
\object{NGC 266} & 62.4  & \citetalias{Tully1988} &  & L1.9 & \citetalias{Ho_etal.1997a} & SB(rs)ab & 8.30  & \citetalias{Doi_etal.2005a},\citetalias{Kormendy_Gebhardt2001},\citetalias{Ho_etal.1997b} & 40.88 & \citetalias{Terashima_Wilson2003} \\
\object{NGC 2787} & 7.5  & \citetalias{Tonry_etal.2001} &  & L1.9 & \citetalias{Ho_etal.1997a} & SB(r)0+ & 7.64 & \citetalias{Sarzi_etal.2001},\citetalias{Heckman_etal.1980} & 38.79 & \citetalias{Ho_etal.2001} \\
\object{NGC 3031} (M81) & 3.6  & \citetalias{Freedman_etal.1994} &  & S1.5 & \citetalias{Ho_etal.1997a} & SA(s)ab & 7.8 & \citetalias{Devereux_etal.2003} & 39.38 & \citetalias{Ho_etal.2001} \\
\object{NGC 3147} & 40.9  & \citetalias{Tonry_etal.2001} &  & S2 & \citetalias{Ho_etal.1997a} & SA(rs)bc & 8.64 & \citetalias{McElroy1995},\citetalias{Tremaine+2002} & 41.87 & \citetalias{Terashima_Wilson2003} \\
\object{NGC 3169} & 19.7  & \citetalias{Tonry_etal.2001} &  & L2 & \citetalias{Ho_etal.1997a} & SA(s)a pec & 7.8 & \citetalias{Eracleous_etal.2010} & 41.05 & \citetalias{Terashima_Wilson2003} \\
\object{NGC 3226} & 23.6  & \citetalias{Tonry_etal.2001} & 3 & L1.9 & \citetalias{Ho_etal.1997a} & E2: pec & 8.1 & \citetalias{Eracleous_etal.2010} & 39.99 & \citetalias{Gonzalez-Martin_etal.2006} \\
\object{NGC 3718} & 17.0  & \citetalias{Tully1988} &  & L1.9 & \citetalias{Ho_etal.1997a} & SB(s)a pec & 7.85 & \citetalias{Ho2007},\citetalias{Tremaine+2002} & 40.44 & \citetalias{Satyapal_etal.2005} \\
\object{NGC 4143} & 15.9  & \citetalias{Tonry_etal.2001} &  & L1.9 & \citetalias{Ho_etal.1997a} & SAB(s)0 & 8.35 & \citetalias{Pellegrini2010} & 40.03 & \citetalias{Terashima_Wilson2003} \\
\object{NGC 4258} & 7.3  & \citetalias{Herrnstein_etal.1999} &  & S1.9 & \citetalias{Ho_etal.1997a} & SAB(s)bc & 7.6 & \citetalias{Herrnstein_etal.1999} & 40.89 & \citetalias{Reynolds_etal.2009} \\
\object{NGC 4278} & 16.1  & \citetalias{Tonry_etal.2001} & 3  & L1.9 & \citetalias{Ho_etal.1997a} & E1+ & 8.53 & \citetalias{Pellegrini2010} & 39.64 & \citetalias{Ho_etal.2001} \\
\object{NGC 4374} (M 84) & 18.4  & \citetalias{Tonry_etal.2001} & 4  & L2 & \citetalias{Ho_etal.1997a} & E1 & 9.2 & \citetalias{Bower_etal.1998} & 39.50 & \citetalias{Gonzalez-Martin_etal.2006} \\
\object{NGC 4565} & 17.5  & \citetalias{Tonry_etal.2001} &  & S1.9 & \citetalias{Ho_etal.1997a} & SA(s)b?spin & 7.46 & \citetalias{Ho2007},\citetalias{Tremaine+2002} & 39.56 & \citetalias{Terashima_Wilson2003} \\
\object{NGC 4579} (M 58) & 19.1  & \citetalias{Solanes_etal.2002} &  & S1.9/L1.9 & \citetalias{Ho_etal.1997a} & SAB(rs)b & 7.8 & \citetalias{Eracleous_etal.2010} & 41.15 & \citetalias{Gonzalez-Martin_etal.2006} \\
\object{NGC 4772} & 16.3  & \citetalias{Tully1988} &  & L1.9 & \citetalias{Ho_etal.1997a} & SA(s)a & 7.70  & \citetalias{Doi_etal.2005b},\citetalias{Kormendy_Gebhardt2001},\citetalias{Ho_etal.1997b} & 39.30 & \citetalias{Halderson_etal.2001} \\
\object{NGC 5866} & 15.3  & \citetalias{Tonry_etal.2001} &  & T2 & \citetalias{Ho_etal.1997a} & SA0 + spin & 7.84 & \citetalias{Pellegrini2010} & 38.60 & \citetalias{Gonzalez-Martin_etal.2006} \\
\object{NGC 6500} & 39.7  & \citetalias{Tully1988} &  & L2 & \citetalias{Ho_etal.1997a} & SAab: & 8.3 & \citetalias{Eracleous_etal.2010} & 39.73 & \citetalias{Terashima_Wilson2003} \\\tableline
\end{tabular}
\end{center}
\end{table*}
\begin{table*}
\caption{Nearby Early-type Galaxy Sample\label{table2}}
\begin{center}
\begin{tabular}{lrccccccccc} \tableline\tableline
\multicolumn{1}{c}{Name} & \multicolumn{1}{c}{$D$} & Ref. & Dust & Line & Ref. & Morph. & $\log{M_\mathrm{BH}}$ & Ref. & $L_\mathrm{X}$ & Ref. \\
\multicolumn{1}{c}{} & \multicolumn{1}{c}{(Mpc)} &  &  & Emission &  &  & ($M_\mathrm{\sun}$) &  & (erg s$^{-1}$) &  \\
\multicolumn{1}{c}{(1)} & \multicolumn{1}{c}{(2)} & (3) & (4) & (5) & (6) & (7) & (8) & (9) & (10) & (11) \\\tableline
\object{NGC 0404} & 3.3  & \citetalias{Tonry_etal.2001} & 3 & L2 & \citetalias{Ho_etal.1997a} & S03(0) & 5.39 & \citetalias{Pellegrini2010} & 37.02 & \citetalias{Eracleous_etal.2002} \\
\object{NGC 2768} & 22.4  & \citetalias{Tonry_etal.2001} & 4 & L2 & \citetalias{Ho_etal.1997a} & E6: & 7.97 & \citetalias{Ho2007},\citetalias{Tremaine+2002} & 40.59 & \citetalias{Komossa_etal.1999} \\
\object{NGC 2974} & 21.5  & \citetalias{Tonry_etal.2001} & 3 & Y & \citetalias{Roberts_etal.1991} & E4 & 8.53 & \citetalias{Pellegrini2010} & 40.32 & \citetalias{Liu_Bregman2005} \\
\object{NGC 3065} & 31.3  & \citetalias{Tully1988} & 4 & Y & \citetalias{Roberts_etal.1991} & S01/2(0) & 8.17 & \citetalias{Pellegrini2010} & 41.32 & \citetalias{Iyomoto_etal.1998} \\
\object{NGC 3156} & 22.4  & \citetalias{Tonry_etal.2001} & 2 & Y & \citetalias{Bettoni_Buson1987} & S0: & 6.61 & \citetalias{Ho2007},\citetalias{Tremaine+2002} & $<40.0$\tablenotemark{a} & \citetalias{OSullivan_etal.2001} \\
\object{NGC 3610} & 21.4  & \citetalias{Tonry_etal.2001} & 3 & Y & \citetalias{Goudfrooij_etal.1994} & E5: & 7.74 & \citetalias{Bender_etal.1994},\citetalias{Tremaine+2002} & 38.79 & \citetalias{Desroches_Ho2009} \\
\object{NGC 4125} & 23.9  & \citetalias{Tonry_etal.2001} & 4 & T2 & \citetalias{Ho_etal.1997a} & E6 pec & 8.45 & \citetalias{Pellegrini2010} & 38.94 & \citetalias{Gonzalez-Martin_etal.2006} \\
\object{NGC 4168} & 30.9  & \citetalias{Tonry_etal.2001} & 2 & S1.9 & \citetalias{Ho_etal.1997a} & E2 & 8.09 & \citetalias{Pellegrini2010} & $<38.24$ & \citetalias{Panessa_etal.2006} \\
\object{NGC 4233} & 33.9  & \citetalias{Tonry_etal.2001} & 4 & Y & \citetalias{Kollatschny_Fricke1989} & S0 & 8.21 & \citetalias{Bernardi_etal.2002},\citetalias{Tremaine+2002} & $<39.22$\tablenotemark{a} & \citetalias{OSullivan_etal.2001} \\
\object{NGC 4494} & 17.1  & \citetalias{Tonry_etal.2001} & 4 & L2:: & \citetalias{Ho_etal.1997a} & E1+ & 7.74 & \citetalias{Pellegrini2010} & 39.12 & \citetalias{Gonzalez-Martin_etal.2006} \\
\object{NGC 4552} (M 89) & 15.3  & \citetalias{Tonry_etal.2001} & 4 & T2: & \citetalias{Ho_etal.1997a} & E & 8.63 & \citetalias{Pellegrini2010} & 39.49 & \citetalias{Gonzalez-Martin_etal.2006} \\
\object{NGC 4589} & 22.0  & \citetalias{Tonry_etal.2001} & 2 & L2 & \citetalias{Ho_etal.1997a} & E2 & 8.43 & \citetalias{Pellegrini2010} & 40.36\tablenotemark{a} & \citetalias{OSullivan_etal.2001} \\
\object{NGC 4697} & 11.7  & \citetalias{Tonry_etal.2001} & 4 & Y & \citetalias{Gonzalez1993} & E6 & 8.28 & \citetalias{Pellegrini2010} & 38.41 & \citetalias{Soria_etal.2006} \\
\object{NGC 4742} & 15.5  & \citetalias{Tonry_etal.2001} & 3 & Y & \citetalias{Roberts_etal.1991} & E4: & 7.18 & \citetalias{Gultekin_etal.2009} & $<39.80$\tablenotemark{a} & \citetalias{OSullivan_etal.2001} \\
\object{NGC 5077} & 40.6  & \citetalias{Tully1988} & 2 & L1.9 & \citetalias{Ho_etal.1997a} & E3+ & 8.90 & \citetalias{Gebhardt_etal.2003} & 40.48\tablenotemark{a} & \citetalias{OSullivan_etal.2001} \\
\object{NGC 5173} & 38.0  & \citetalias{Tully1988} & 3 & Y & \citetalias{Bettoni_Buson1987} & E0: & 6.96 & \citetalias{Wegner_etal.2003},\citetalias{Tremaine+2002} & $<40.36$\tablenotemark{a} & \citetalias{OSullivan_etal.2001} \\
\object{NGC 5283} & 41.4  & \citetalias{Tully1988} & 3 & S2 & \citetalias{Huchra_Burg1992} & S0: & 7.71 & \citetalias{Pellegrini2010} & 41.91 & \citetalias{Ghosh_etal.2007} \\
\object{NGC 5322} & 31.2  & \citetalias{Tonry_etal.2001} & 4 & L2:: & \citetalias{Ho_etal.1997a} & E3+ & 8.49 & \citetalias{Pellegrini2010} & 40.26 & \citetalias{Liu_Bregman2005} \\
\object{NGC 5812} & 26.9  & \citetalias{Tonry_etal.2001} & 4 & Y & \citetalias{Gonzalez1993} & E0 & 8.05 & \citetalias{Kuntschner_etal.2001},\citetalias{Tremaine+2002} & $<40.32$\tablenotemark{a} & \citetalias{OSullivan_etal.2001} \\
\object{NGC 5813} & 32.2  & \citetalias{Tonry_etal.2001} & 4 & L2: & \citetalias{Ho_etal.1997a} & E1+ & 8.52 & \citetalias{Pellegrini2010} & 38.51 & \citetalias{Pellegrini2005} \\
\object{NGC 5846} & 24.9  & \citetalias{Tonry_etal.2001} & 3 & T2: & \citetalias{Ho_etal.1997a} & E0+ & 8.54 & \citetalias{Pellegrini2010} & 39.65 & \citetalias{Gonzalez-Martin_etal.2006} \\\tableline
\end{tabular}
\end{center}
\tablecomments{Col.~(1)~source name; Col.~(2)~distance; Col.~(3)~reference for distance; Col.~(4)~dust level, 1:~filamentary low, 2:~filamentary medium, 3:~filamentary high, 4:~dusty disk \citep{Tran_etal.2001}; Col.~(5)~classification of the nuclear spectrum as described in \citet{Ho_etal.1997b} and \citet{Tran_etal.2001}: S=Seyfert nucleus, L=LINER, T=transition object, Y=emission lines detected.  The number attached to the class letter designates the type (1.0, 1.2, 1.5, 1.8, 1.9, and 2); quality ratings are given by : and :: for uncertain and highly uncertain classifications, respectively; Col.~(6)~reference of classification; Col.~(7)~Hubble type in \citet{Ho_etal.1997b} or \citet{Roberts_etal.1991}; Col.~(8)~black hole mass; Col.~(9)~ reference of black hole mass; Col.~(10)~X-ray luminosity at 2--10~keV; Col.~(11)~reference of X-ray luminosity.}
\tablenotemark{a}{``pseudo-bolometric X-ray luminosity'' converted from the data of the {\it Einstein} IPC or {\it ROSAT} PSPC \citep[see][]{OSullivan_etal.2001}, an overestimated value as 2--10-keV X-ray luminosity.}
\tablerefs{
(\citetalias{Tully1988})~\citealt{Tully1988}, 
(\citetalias{Tonry_etal.2001})~\citealt{Tonry_etal.2001}, 
(\citetalias{Freedman_etal.1994})~\citealt{Freedman_etal.1994}, 
(\citetalias{Herrnstein_etal.1999})~\citealt{Herrnstein_etal.1999}, 
(\citetalias{Solanes_etal.2002})~\citealt{Solanes_etal.2002}, 
(\citetalias{Ho_etal.1997a})~\citealt{Ho_etal.1997a}, 
(\citetalias{Roberts_etal.1991})~\citealt{Roberts_etal.1991}, 
(\citetalias{Bettoni_Buson1987})~\citealt{Bettoni_Buson1987}, 
(\citetalias{Goudfrooij_etal.1994})~\citealt{Goudfrooij_etal.1994}, 
(\citetalias{Kollatschny_Fricke1989})~\citealt{Kollatschny_Fricke1989}, 
(\citetalias{Gonzalez1993})~\citealt{Gonzalez1993}, 
(\citetalias{Huchra_Burg1992})~\citealt{Huchra_Burg1992}, 
(\citetalias{Doi_etal.2005a})~\citealt{Doi_etal.2005a}, 
(\citetalias{Kormendy_Gebhardt2001})~\citealt{Kormendy_Gebhardt2001}, 
(\citetalias{Ho_etal.1997b})~\citealt{Ho_etal.1997b}, 
(\citetalias{Sarzi_etal.2001})~\citealt{Sarzi_etal.2001}, 
(\citetalias{Heckman_etal.1980})~\citealt{Heckman_etal.1980}, 
(\citetalias{Devereux_etal.2003})~\citealt{Devereux_etal.2003}, 
(\citetalias{McElroy1995})~\citealt{McElroy1995}, 
(\citetalias{Eracleous_etal.2010})~\citealt{Eracleous_etal.2010}, 
(\citetalias{Ho2007})~\citealt{Ho2007}, 
(\citetalias{Tremaine+2002})~\citealt{Tremaine+2002}, 
(\citetalias{Pellegrini2010})~\citealt{Pellegrini2010}, 
(\citetalias{Bower_etal.1998})~\citealt{Bower_etal.1998}, 
(\citetalias{Doi_etal.2005b})~\citealt{Doi_etal.2005b}, 
(\citetalias{Bender_etal.1994})~\citealt{Bender_etal.1994}, 
(\citetalias{Bernardi_etal.2002})~\citealt{Bernardi_etal.2002}, 
(\citetalias{Gultekin_etal.2009})~\citealt{Gultekin_etal.2009}, 
(\citetalias{Gebhardt_etal.2003})~\citealt{Gebhardt_etal.2003}, 
(\citetalias{Wegner_etal.2003})~\citealt{Wegner_etal.2003}, 
(\citetalias{Kuntschner_etal.2001})~\citealt{Kuntschner_etal.2001}, 
(\citetalias{Terashima_Wilson2003})~\citealt{Terashima_Wilson2003}, 
(\citetalias{Ho_etal.2001})~\citealt{Ho_etal.2001}, 
(\citetalias{Gonzalez-Martin_etal.2006})~\citealt{Gonzalez-Martin_etal.2006}, 
(\citetalias{Satyapal_etal.2005})~\citealt{Satyapal_etal.2005}, 
(\citetalias{Reynolds_etal.2009})~\citealt{Reynolds_etal.2009}, 
(\citetalias{Halderson_etal.2001})~\citealt{Halderson_etal.2001}, 
(\citetalias{Eracleous_etal.2002})~\citealt{Eracleous_etal.2002}, 
(\citetalias{Komossa_etal.1999})~\citealt{Komossa_etal.1999}, 
(\citetalias{Liu_Bregman2005})~\citealt{Liu_Bregman2005}, 
(\citetalias{Iyomoto_etal.1998})~\citealt{Iyomoto_etal.1998}, 
(\citetalias{OSullivan_etal.2001})~\citealt{OSullivan_etal.2001}, 
(\citetalias{Desroches_Ho2009})~\citealt{Desroches_Ho2009}, 
(\citetalias{Panessa_etal.2006})~\citealt{Panessa_etal.2006}, 
(\citetalias{Soria_etal.2006})~\citealt{Soria_etal.2006}, 
(\citetalias{Ghosh_etal.2007})~\citealt{Ghosh_etal.2007}, 
(\citetalias{Pellegrini2005})~\citealt{Pellegrini2005}
}
\end{table*}

In addition, we conducted new NMA observations for the subsample of five LLAGNs at $\lambda3$mm and $\lambda2$mm by using the D-array configuration on 2004 April~24--26; the HPBWs of the synthesized beams were typically $\sim 7 \arcsec$ and $5 \arcsec$, respectively.  In an observation day, we performed observations for each source quasi-simultaneously at $\lambda\lambda3$ and 2~mm.  Visibility data were obtained with double-sided-band receiving systems at center frequencies of 89.725 and 101.725 GHz for $\lambda3$~mm and of 134.969 and 146.969 GHz for $\lambda2$~mm, which were Doppler-tracked.  These observing frequencies with a bandwidth of 1~GHz each (4~GHz in total) were determined to avoid contamination due to the several significant emission lines of molecular gas in typical extra galaxies.  The sky conditions during the observations were good.  In the double-sided band, the typical system noise temperature was about 150~K and 200~K at $\lambda3$mm and $\lambda2$mm, respectively.  Amplitude, phase, bandpass, and flux scaling calibrations were performed as mentioned earlier.     

We conducted new observations for the 19~early-type galaxies~(NGC~4168 and NGC~4552 were excluded) at $\lambda3$mm using the NMA with the D configuration.  The observations were conducted between 2002 November~27 and 2003 May~25 at the center frequencies of 89.725 and 101.725 GHz.  Such frequency bands can avoid possible contamination due to the several significant line emissions of the sources in the sample.  A system noise temperature, $T_\mathrm{sys}$, was typically about 150~K in the double-sided bands.  Amplitude, phase, bandpass, and flux scaling calibrations were performed as mentioned above.

\begin{table*}
\caption{Observation Results of Nearby LLAGN Sample\label{table3}}
\begin{center}
\begin{tabular}{l|lcccccc} \hline\hline 
\multicolumn{1}{c}{Name} & \multicolumn{1}{c}{Date} & $S_\mathrm{3mm}$ & $\sigma^\mathrm{rms}_\mathrm{3mm}$ & $\nu_\mathrm{3mm}$ & $S_\mathrm{2mm}$ & $\sigma^\mathrm{rms}_\mathrm{2mm}$ & $\nu_\mathrm{2mm}$ \\
\multicolumn{1}{c}{} & \multicolumn{1}{c}{} & (mJy) & (mJy beam$^{-1}$) & (GHz) & (mJy) & (mJy beam$^{-1}$) & (GHz) \\
\multicolumn{1}{c}{(1)} & \multicolumn{1}{c}{(2)} & (3) & (4) & (5) & (6) & (7) & (8) \\\hline
\multicolumn{1}{l}{NGC 266} & \multicolumn{1}{l}{2003 Apr 26} & $<9.2$ & 3.1  & 95.725 &  &  &  \\
\multicolumn{1}{l}{} & \multicolumn{1}{l}{2003 Apr 28} & $<6.3$ & 2.1  & 95.725 &  &  &  \\
\multicolumn{1}{l}{} & \multicolumn{1}{l}{2003 May 13} & $<5.2$ & 1.7  & 95.725 &  &  &  \\
\multicolumn{1}{l}{} & \multicolumn{1}{l}{2003 May 14} & $<7.5$ & 2.5  & 95.725 &  &  &  \\
\multicolumn{1}{l}{} & \multicolumn{1}{l}{2003 May 25} & $<7.6$ & 2.5  & 95.725 &  &  &  \\
\multicolumn{1}{l}{} & \multicolumn{1}{l}{(all data)} & $<2.7$ & 0.9  & 95.725 &  &  &  \\
\multicolumn{1}{l}{NGC 2787} & \multicolumn{1}{l}{2002 Nov 28} & $22.5 \pm 5.2$ & 2.2  & 95.725 &  &  &  \\
\multicolumn{1}{l}{} & \multicolumn{1}{l}{2002 Nov 29} & $16.2 \pm 6.4$ & 3.5  & 95.725 &  &  &  \\
\multicolumn{1}{l}{} & \multicolumn{1}{l}{2002 Dec 14} & $13.1 \pm 2.2$ & 1.0  & 95.725 &  &  &  \\
\multicolumn{1}{l}{} & \multicolumn{1}{l}{2004 Apr 25} & $15.4 \pm 2.9$ & 1.3  & 95.725 &  &  &  \\
\multicolumn{1}{l}{NGC 3031} & \multicolumn{1}{l}{2004 Apr 24} & $81.8 \pm 9.4$ & 2.4  & 95.725 & $109.9 \pm 16.7$ & 2.9  & 140.969 \\
\multicolumn{1}{l}{NGC 3147} & \multicolumn{1}{l}{2002 Dec 12} & $<3.943$ & 1.3  & 95.725 &  &  &  \\
\multicolumn{1}{l}{} & \multicolumn{1}{l}{2002 Dec 14} & $4.7 \pm 2.2$ & 1.3  & 95.725 &  &  &  \\
\multicolumn{1}{l}{NGC 3169} & \multicolumn{1}{l}{2003 Apr 24} & $<7.0$ & 2.3  & 95.725 &  &  &  \\
\multicolumn{1}{l}{} & \multicolumn{1}{l}{2003 May 25} & $10.1 \pm 5.4$ & 2.7  & 95.725 &  &  &  \\
\multicolumn{1}{l}{NGC 3226} & \multicolumn{1}{l}{2002 Dec 24} & $9.2 \pm 2.6$ & 1.4 & 95.725 &  &  &  \\
\multicolumn{1}{l}{} & \multicolumn{1}{l}{2002 Dec 25} & $8.1 \pm 2.6$ & 1.3  & 95.725 &  &  &  \\
\multicolumn{1}{l}{NGC 3718} & \multicolumn{1}{l}{2003 Apr 22} & $<5.5$ & 1.8  & 95.725 &  &  &  \\
\multicolumn{1}{l}{} & \multicolumn{1}{l}{2003 Apr 24} & $<9.6$ & 3.2  & 95.725 &  &  &  \\
\multicolumn{1}{l}{} & \multicolumn{1}{l}{2003 May 25} & $20.0 \pm 6.0$ & 2.4  & 95.725 &  &  &  \\
\multicolumn{1}{l}{} & \multicolumn{1}{l}{2004 Apr 25} & $15.8 \pm 4.2$ & 2.1  & 95.725 &  &  &  \\
\multicolumn{1}{l}{NGC 4143} & \multicolumn{1}{l}{2003 May 09} & $7.6 \pm 3.2$ & 1.4  & 95.725 &  &  &  \\
\multicolumn{1}{l}{} & \multicolumn{1}{l}{2003 May 10} & $<7.3$ & 2.4  & 95.725 &  &  &  \\
\multicolumn{1}{l}{} & \multicolumn{1}{l}{2003 May 12} & $<10.8$ & 3.6  & 95.725 &  &  &  \\
\multicolumn{1}{l}{NGC 4258} & \multicolumn{1}{l}{2001 Dec 21} & $10.0 \pm 1.7$ & 0.9  & 100.777 &  &  &  \\
\multicolumn{1}{l}{} & \multicolumn{1}{l}{2002 Mar 21} & $18.1 \pm 4.0$ & 3.0  & 100.777 &  &  &  \\
\multicolumn{1}{l}{} & \multicolumn{1}{l}{2003 Dec 22} & $5.2 \pm 1.2$ & 0.9  & 100.777 &  &  &  \\
\multicolumn{1}{l}{} & \multicolumn{1}{l}{2004 Jan 14} & $<5.9$ & 2.0  & 100.777 &  &  &  \\
\multicolumn{1}{l}{} & \multicolumn{1}{l}{2004 Mar 31} & $6.2 \pm 2.0$ & 1.8  & 100.777 &  &  &  \\
\multicolumn{1}{l}{} & \multicolumn{1}{l}{2004 Apr 01} & $<8.3$ & 2.8  & 100.777 &  &  &  \\
\multicolumn{1}{l}{} & \multicolumn{1}{l}{2005 Apr 05} & $4.6 \pm 1.8$ & 0.8 & 100.777 &  &  &  \\
\multicolumn{1}{l}{} & \multicolumn{1}{l}{2005 Apr 08} & $8.0 \pm 3.3$ & 1.5 & 100.777 &  &  &  \\
\multicolumn{1}{l}{} & \multicolumn{1}{l}{2005 May 13} & $8.4 \pm 2.0$ & 0.9 & 95.729 &  &  &  \\
\multicolumn{1}{l}{} & \multicolumn{1}{l}{2005 May 14} & $10.4 \pm 4.0$ & 1.9 & 95.729 &  &  &  \\
\multicolumn{1}{l}{} & \multicolumn{1}{l}{2005 May 15} & $7.0 \pm 2.1$ & 0.9 & 95.729 &  &  &  \\
\multicolumn{1}{l}{} & \multicolumn{1}{l}{2006 Mar 25} & $<9.4$ & 3.1  & 100.777 &  &  &  \\
\multicolumn{1}{l}{} & \multicolumn{1}{l}{2006 Mar 29} & $<9.4$ & 3.1  & 100.777 &  &  &  \\
\multicolumn{1}{l}{} & \multicolumn{1}{l}{2006 Mar 31} & $<7.1$ & 2.4  & 100.777 &  &  &  \\
\multicolumn{1}{l}{} & \multicolumn{1}{l}{2006 Apr 01} & $<8.6$ & 2.9  & 100.777 &  &  &  \\
\multicolumn{1}{l}{NGC 4278} & \multicolumn{1}{l}{2003 May 10} & $57.0 \pm 9.1$ & 3.1  & 95.725 &  &  &  \\
\multicolumn{1}{l}{} & \multicolumn{1}{l}{2004 Apr 24} & $87.2 \pm 14.0$ & 5.0  & 95.725 & $64.0 \pm 11.5$ & 5.0 & 140.969 \\
\multicolumn{1}{l}{} & \multicolumn{1}{l}{2004 Apr 25} & $55.6 \pm 9.2$ & 3.8  & 95.725 & $42.5 \pm 9.8$ & 7.5 & 146.969 \\
\multicolumn{1}{l}{NGC 4374} & \multicolumn{1}{l}{2003 May 10} & $146.7 \pm 22.6$ & 5.0  & 95.725 &  &  &  \\
\multicolumn{1}{l}{} & \multicolumn{1}{l}{2004 Apr 24} & $169.6 \pm 26.4$ & 7.1  & 95.725 & $171.7 \pm 26.4$ & 5.8 & 140.969 \\
\multicolumn{1}{l}{NGC 4565} & \multicolumn{1}{l}{2003 Apr 25} & $<14.9$ & 5.0  & 95.725 &  &  &  \\
\multicolumn{1}{l}{} & \multicolumn{1}{l}{2003 May 10} & $<5.5$ & 1.8  & 95.725 &  &  &  \\
\multicolumn{1}{l}{} & \multicolumn{1}{l}{2003 May 23} & $<10.1$ & 3.4  & 95.725 &  &  &  \\
\multicolumn{1}{l}{} & \multicolumn{1}{l}{2003 May 25} & $<9.6$ & 3.2  & 95.725 &  &  &  \\
\multicolumn{1}{l}{} & \multicolumn{1}{l}{(all data)} & $<4.7$ & 1.6  & 95.725 &  &  &  \\
\multicolumn{1}{l}{NGC 4579} & \multicolumn{1}{l}{2003 Dec 25} & $28.6 \pm 7.8$ & 2.7  & 95.725 &  &  &  \\
\multicolumn{1}{l}{} & \multicolumn{1}{l}{2004 Apr 26} & $26.8 \pm 4.6$ & 1.9  & 95.725 & $26.8 \pm 9.1$ & 4.9 & 134.969 \\
\multicolumn{1}{l}{NGC 4772} & \multicolumn{1}{l}{2002 Dec 24} & $5.2 \pm 2.9$ & 1.4  & 95.725 &  &  &  \\
\multicolumn{1}{l}{} & \multicolumn{1}{l}{2002 Dec 25} & $8.6 \pm 3.7$ & 1.7  & 95.725 &  &  &  \\
\multicolumn{1}{l}{NGC 5866} & \multicolumn{1}{l}{2003 Apr 28} & $<6.6$ & 2.2  & 95.725 &  &  &  \\
\multicolumn{1}{l}{} & \multicolumn{1}{l}{2003 May 09} & $<3.7$ & 1.2  & 95.725 &  &  &  \\
\multicolumn{1}{l}{} & \multicolumn{1}{l}{2003 May 12} & $<11.9$ & 4.0  & 95.725 &  &  &  \\
\multicolumn{1}{l}{} & \multicolumn{1}{l}{2003 May 24} & $<11.6$ & 3.9  & 95.725 &  &  &  \\
\multicolumn{1}{l}{} & \multicolumn{1}{l}{2003 May 25} & $<4.7$ & 1.6  & 95.725 &  &  &  \\
\multicolumn{1}{l}{} & \multicolumn{1}{l}{(all data)} & $<2.9$ & 1.0  & 95.725 &  &  &  \\
\multicolumn{1}{l}{NGC 6500} & \multicolumn{1}{l}{2003 Apr 24} & $37.0 \pm 18.1$ & 8.2  & 95.725 &  &  &  \\
\multicolumn{1}{l}{} & \multicolumn{1}{l}{2003 Apr 28} & $43.9 \pm 7.4$ & 3.4  & 95.725 &  &  &  \\
\multicolumn{1}{l}{} & \multicolumn{1}{l}{2004 Apr 26} & $48.3 \pm 8.1$ & 3.5  & 95.725 & $52.6 \pm 10.3$ & 4.0 & 140.969 \\
\multicolumn{1}{l}{} & \multicolumn{1}{l}{2004 Apr 27} & $46.5 \pm 12.7$ & 6.9  & 95.725 &  &  &  \\\hline
\end{tabular}
\end{center}
\tablecomments{Col.~(1)~source name; Col.~(2)~observation date; Col.~(3)~flux density at 3~mm; Col.~(4)~r.m.s.~of image noise; Col.~(5)~observing frequency for 3~mm; Col.~(6)~flux density at 2~mm, Col.~(7)~r.m.s.~of image noise; Col.~(8)~observing frequency for 2~mm.}
\end{table*}
\begin{table}
\caption{Observation Results of Nearby Early-type Galaxy Sample.\label{table4}}
\begin{center}
\begin{tabular}{llccc} \hline\hline
\multicolumn{1}{c}{Name} & \multicolumn{1}{c}{Date} & $S_\mathrm{3mm}$ & $\sigma^\mathrm{rms}_\mathrm{3mm}$ & $\nu_\mathrm{3mm}$ \\
\multicolumn{1}{c}{} & \multicolumn{1}{c}{} & (mJy) & (mJy beam$^{-1}$) & (GHz) \\
\multicolumn{1}{c}{(1)} & \multicolumn{1}{c}{(2)} & (3) & (4) & (5) \\\hline
NGC 0404 & 2003 May 13 & $<8.4$ & 2.8  & 95.725 \\
NGC 2768 & 2002 Dec 24 & $<5.3$ & 1.8  & 95.725 \\
NGC 2974 & 2003 May 24 & $<9.4$ & 3.1  & 95.725 \\
NGC 3065 & 2003 Apr 22 & $<5.3$ & 1.8  & 95.725 \\
NGC 3156 & 2003 May 23 & $<10.0$ & 3.3  & 95.725 \\
NGC 3610 & 2003 May 09 & $<6.5$ & 2.2  & 95.725 \\
NGC 4125 & 2002 Dec 12 & $<5.6$ & 1.9  & 95.725 \\
NGC 4168 & 2002 Dec 13 & $6.0 \pm 2.5$ & 1.3  & 95.725 \\
NGC 4233 & 2003 May 23 & $<6.9$ & 2.3  & 95.725 \\
NGC 4494 & 2003 May 24 & $<7.0$ & 2.3  & 95.725 \\
NGC 4552 & 2002 Dec 24 & $12.2 \pm 4.1$ & 2.6  & 95.725 \\
NGC 4589 & 2002 Dec 14 & $7.4 \pm 3.0$ & 2.0  & 95.725 \\
NGC 4697 & 2003 May 11 & $<6.7$ & 2.2  & 95.725 \\
NGC 4742 & 2003 May 11 & $<15.8$ & 5.3  & 95.725 \\
NGC 5077 & 2002 Nov 27 & $657.6 \pm 113.2$ & 47.0  & 95.725 \\
NGC 5173 & 2003 May 11 & $<11.0$ & 3.7  & 95.725 \\
NGC 5283 & 2003 May 25 & $<7.5$ & 2.5  & 95.725 \\
NGC 5322 & 2003 May 23 & $21.1 \pm 7.4$ & 3.0  & 89.725 \\
NGC 5812 & 2003 May 24 & $<7.4$ & 2.5  & 95.725 \\
NGC 5813 & 2003 May 24 & $<8.2$ & 2.7  & 95.725 \\
NGC 5846 & 2003 May 23 & $9.0 \pm 4.3$ & 2.5  & 95.725 \\\hline
\end{tabular}
\end{center}
\tablecomments{Col.~(1)~source name; Col.~(2)~observation date; Col.~(3)~flux density; Col.~(4)~r.m.s.~of image noise; Col.~(5)~observing frequency.}
\end{table}

The data were reduced using the {\tt UVPROC-II} package \citep{Tsutsumi_etal.1997} developed at the Nobeyama Radio Observatory~(NRO) by standard procedures including bad data flagging, baseline correction, opacity correction, bandpass calibration, and gain calibration.  No signature of emission lines was found in the data for either sources.  To achieve higher sensitivity, visibilities of both the sidebands were combined with the same weight, which resulted in a center frequency of 95.725 GHz for the lower and upper sideband frequencies of 89.725 and 101.725 GHz, respectively.  For the data of NGC~4278 on 2004 April~25 and NGC~4579 on 2004 April~26, only one sideband was used for $\lambda=2$~mm because of poor quality.  Each daily image was individually created in natural weighting and was deconvolved using the {\tt AIPS} software developed at the National Radio Astronomy Observatory (NRAO).  The results for visibilities combined over all epochs have been published for LLAGNs \citep{Doi_etal.2005b}.  The rms~of the noise in the images was estimated from statistics on off-source blank sky with the {\tt AIPS} task {\tt IMEAN}.  If an emission with a peak intensity of more than three times the rms of the image noise was found at the galactic center, source identification and flux density measurements were performed in the image domain by elliptical Gaussian profile fitting with the {\tt JMFIT} task.  We derived the total errors in the flux measurements from the root sum square of the errors in Gaussian fitting (including thermal noise) and flux scaling~(15\%).  For negative detection cases, an upper limit of flux density was determined to be three-times the rms~of image noise.

\section{RESULTS}\label{section:result}

The results of the multi-epoch flux measurements for LLAGNs and the multi-frequency measurements for their subsample are shown in Table~\ref{table3}.  For sources undetected in all epochs, the flux measurements on the image created from visibilities combined over all epochs are also listed.  The results of flux measurements for the early-type galaxies are shown in Table~\ref{table4}.  Spectral plots (from our measurements as well as the literature) for the subsample of five LLAGNs observed at multiple frequencies and the four early-type galaxies newly detected at 100~GHz are shown in Figure~\ref{figure1}.  (The spectral plots for NGC~4168 and NGC~4522 have been reported in Figure~1 in \citealt{Doi_etal.2005b}.)

\subsection{Estimation of contamination}
All emissions detected in our NMA observations were point-like with a typical beam size of 7~$\arcsec$.  They originated from the central regions of galaxies; the beam size of 7~$\arcsec$ corresponds to radii of 60~pc or 700~pc at the nearest (NGC~3031) or most distant (NGC~6500) source, respectively.   

To estimate the contribution of nonthermal emissions from supernova remnants and that of extended jet components to the NMA beam, a spectral upper limit was determined by measurements at a lower frequency with a beam much larger than the NMA beam.  The NRAO VLA Sky Survey~(NVSS) conducted with the VLA D-array configuration at 1.4~GHz \citep[beam size of $\sim45 \arcsec$;][]{Condon_etal.1998} served as the upper limit of the extended synchrotron contribution to higher frequencies by assuming the spectral index $\alpha=-0.5$ ($S_\nu \propto \nu^{\alpha}$).  To estimate the contribution of dust emission to the NMA beam, a spectral upper limit was determined by measurements at higher frequencies with beams much larger than the NMA beam.  The upper limit of dust emission was determined by far-infrared measurements (such as {\it Spitzer/ISO/IRAS}) via one- or two-temperature dust spectral modeling using modified blackbody emission $S_\nu \propto \nu^{\beta} B(T_\mathrm{D}$), where $B(T_\mathrm{D})$ is the blackbody spectrum at temperature $T_\mathrm{D}$ \citep[e.g.,][and references therein]{Temi_etal.2004}.  Assuming $\beta$ of dust emissivity to be 1 provides a spectrum that depends on $\nu^3$ at the Rayleigh--Jeans regime and the maximal estimation for lower frequencies (almost all observations of external galaxies report $\beta \sim 1.6$--2 \citep[][and references therein]{Temi_etal.2004}).  In all cases, modeling of the spectral component of the dust temperature resulted in $\sim20$--30~K and provided the upper limit of dust contamination at millimeter wavelengths.

\subsection{Individual galaxies}
The properties of continuum spectrum and flux variability for several galaxies are described as follows.  

\paragraph{NGC~2768} Continuum emission was undetected with the upper limit of 5.3~mJy at 100~GHz.  Using the Plateau de Bure Interferometer~(PdBI), \citet{Crocker_etal.2008} reported detection with 1.94~mJy and undetection with $<2.25$~mJy at 115 and 230~GHz in 2005 December and 2006 August, respectively.       

\paragraph{NGC~3031 (M~81)} The highly variable nuclear emission with a slightly inverted spectrum of M~81 is well known \citep{Ho_etal.1999,Sakamoto_etal.2001,Schodel_etal.2007,Markoff_etal.2008}.  Simultaneous observations at 100~GHz and 230~GHz suggested a turnover frequency between these frequencies \citep{Schodel_etal.2007}, which was in good agreement with the data presented by \citet{Reuter&Lesch1996}; however, these data were not acquired simultaneously.  Our simultaneous observations at 100~GHz and 150~GHz showed an inverted spectrum and that the turnover frequency exists at $>150$~GHz in this epoch.

\paragraph{NGC~3718} The emission at 100~GHz was undetected with the upper limit $<5.5$~mJy on 2003 April~22, whereas it was significantly detected to be $20 \pm 6.0$~mJy after one month (Table~\ref{table3}).  On the basis of the VLA and VLBI spectra at lower frequencies \citep{Doi_etal.2005b}, the emission at 100~GHz is presumed to be a variable, very compact nuclear component with a flat spectrum.  \citet{Krips_etal.2005} reported millimeter continuum detections with $9.4\pm0.7$~mJy~beam$^{-1}$ at 3~mm and $10\pm2$~mJy~beam$^{-1}$ at 1~mm on 2000~January and $7\pm1$~mJy~beam$^{-1}$ at 3~mm on 2001~February; they pointed out a spectral index of $\alpha=0.01\pm0.1$ based on these millimeter measurements and VLA flux densities at centimeter bands in literatures, which is consistent with our picture for NGC~3718.       

\paragraph{NGC~4258} VLBI images showed a nuclear jet associated with the maser disk in NGC~4258 with the average flux density of about 3~mJy and varies by $\sim100$\% on timescales of weeks \citep{Herrnstein_etal.1997}.  The measured flux densities at 100~GHz were from $<5.9$~mJy to 18~mJy, indicating significant flux variability, which cannot be attributed to the dust or extended nonthermal components such as ``the anomalous arms'' of kpc-scale jets \citep{vanderKruit_etal.1972,Hyman_etal.2001}.  The overall radio spectra of the nuclear radio component \citep{Doi_etal.2005b} indicate that NGC~4258 has a highly variable, inverted-spectrum radio nucleus.  An extensive study including VLA spectral evolution as well as the flux variability at 100~GHz for the nuclear radio component and its physical interpretation is presented in \citet{Doi_etal.2011}.

\paragraph{NGC~4278} The 100- and 150-GHz emissions found in our NMA observations showed steep spectra in the two epochs of 2004 April, which may be consistent with the spectrum extrapolated from the NVSS flux density at 1.4~GHz and with the simultaneous VLA spectrum \citep{Nagar_etal.2001}.  However, a flux decrease of $\sim35$\% was found at both 100 and 150~GHz in only two days, which suggests a very compact nuclear component.  In addition, \citet{Crocker_etal.2011} reported flux densities of 32~mJy and 65~mJy at 230~GHz on 2005 December~24 and 2007 December~1, respectively, using PdBI.  The flat spectrum at higher frequencies \citep[150--350~GHz;][]{Anton_etal.2004} may support the picture of nuclear origin of millimeter continuum emissions.  The VLBI image revealed a two-sided milliarcsecond-scale structure with symmetric {\tt S}-shaped jets emerging from a flat-spectrum core \citep{Giroletti_etal.2005a}.

\paragraph{NGC~4374 (M~84)} This is a large radio galaxy with a two-sided jet emerging from its compact core.  The upper limit of nonthermal extended emission obtained by the single-dish observation at 1.4~GHz in Figure~\ref{figure1} might be too large a constraint.  The VLA, VLBI, and James Clerk Maxwell telescope flux densities suggested a flat spectrum at centimeter-to-submillimeter wavelengths.  NMA observations at 100--150~GHz were consistent with the flat spectrum and cannot be attributed to the dust emission or extended nonthermal emission.  Hence, NGC~4278 has a compact nuclear component with a flat spectrum at centimeter-to-submillimeter regimes.  

\paragraph{NGC~4579 (M~58)} The emissions observed at 100 and 150~GHz using the NMA showed a significant excess with respect to the estimated upper-limit spectra of the extended nonthermal emission and dust emission.  The VLA and VLBI spectra at lower frequencies suggest that NGC~4579 has a compact nuclear component with a flat spectrum, which probably continues up to at least 150~GHz.  

\paragraph{NGC~6500} The flux variability and the spectral index in our NMA measurements are unclear.  The VLA flux density measured with a much smaller beam size ($\sim0\arcsec.15$) at 15~GHz indicated a steep spectrum of $\alpha \sim -0.5$ up to 100~GHz or higher.  The two-sided pc-scale jets were found in the VLBI image by \citet{Falcke_etal.2000}; the NMA flux densities may originate in the nonthermal emission of the pc-scale jets.  
  
\paragraph{NGC~4589} We cannot rule out that the emission detected using the NMA originates in the dust or extended nonthermal components.  The VLBI image at 5~GHz showed a core-dominated radio structure with pc-scale jets \citep{Nagar_etal.2005}.    
 
\paragraph{NGC~5077} The emissions detected using the NMA together with VLA and VLBI spectra at lower frequencies originate in a compact nuclear component with a variable, flat spectrum.  The VLBI image at 8.4~GHz showed an apparently two-sided radio structure of continuous jets with the extent of $\sim$10~mas \citep{Petrov_etal.2006}.  

\paragraph{NGC~5322} The emissions observed at 100~GHz using the NMA showed a significant excess with respect to the estimated upper-limit spectra of the extended nonthermal and dust emissions.  The NMA flux density together with the VLA and VLBI spectra at lower frequencies suggest a compact nuclear component with a flat or inverted spectrum in NGC~5322.   

\paragraph{NGC~5846} The emissions observed at 100~GHz using the NMA showed a significant excess with respect to the estimated upper-limit spectra of the extended nonthermal and dust emissions.  The NMA flux density together with the VLA and VLBI spectra at lower frequencies suggest a compact nuclear component with a flat spectrum in NGC~5846.  The VLBI images by \citet{Filho_etal.2004} suggested symmetric two-sided jets found at 2.3 and 5~GHz and an inverted-spectrum core found only at 15~GHz.  The overall radio size was $\sim 100$~mas in the north--south direction.  We speculate that the inverted-spectrum core was the emitting source at 100~GHz.

\setcounter{figure}{1}
\begin{figure*}
\epsscale{1.15}
\plotone{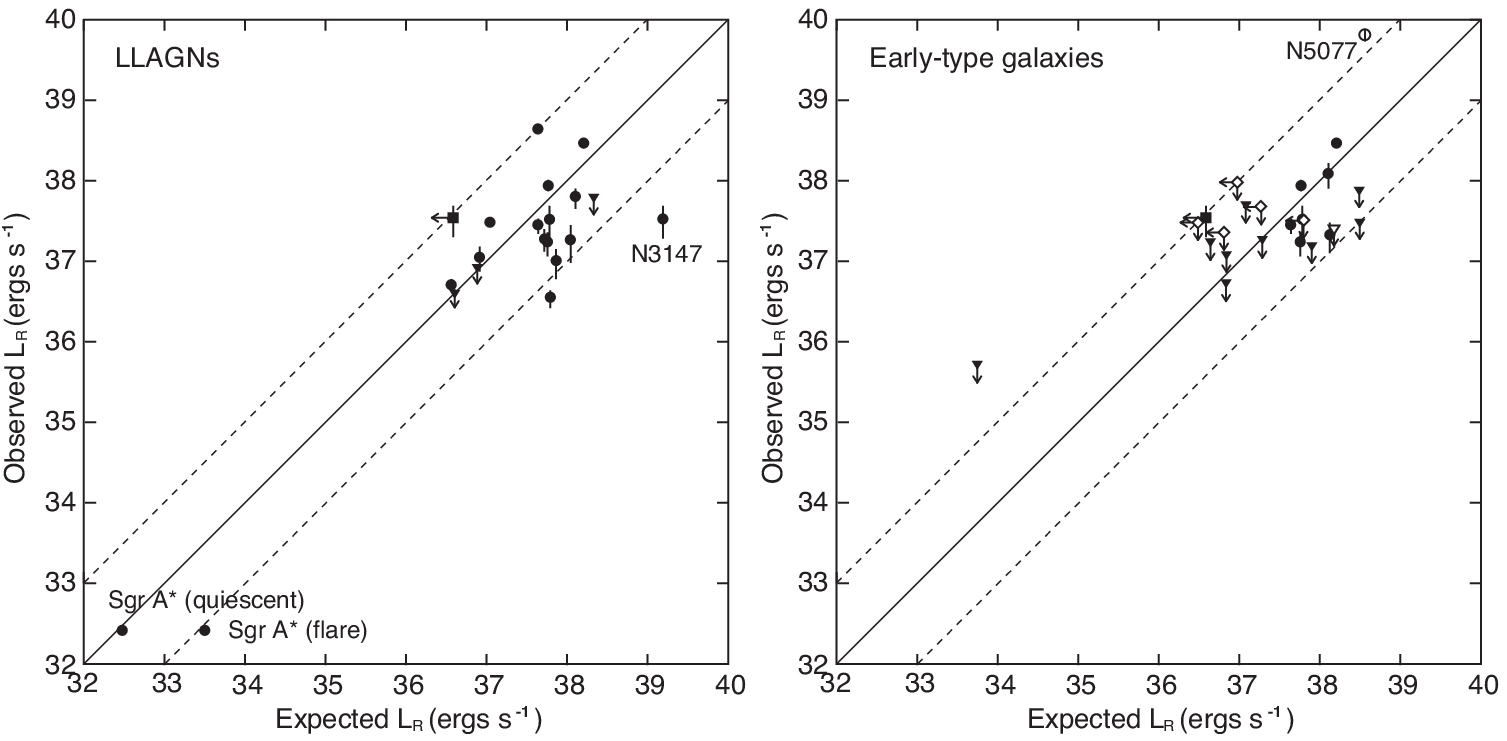}
\figcaption{Comparisons between observed radio luminosity and expected radio luminosity for LLAGNs (left) and early-type galaxies (right).  The expected radio luminosity is calculated by the relation of the fundamental plane of black hole activity \citep{Merloni_etal.2003}.  Solid line is where the observed radio luminosity equals to the fundamental plane; dashed lines show a $\pm$1$\sigma$ scatter from the relation for the sample of \citet{Merloni_etal.2003} (Equation~\ref{eq:fundamentalplane} in \S~\ref{section:discussion}).  The observed radio luminosities are at 5~GHz, converted from 100~GHz assuming a flat spectrum.  Filled symbols are based on X-ray luminosities at 2--10~keV, while opened ones are based on ``pseudo-bolometric X-ray luminosity'' converted from the data of the {\it Einstein} IPC or {\it ROSAT} PSPC \citep[see][]{OSullivan_etal.2001}, which may be an overestimated value for 2--10~keV X-ray luminosity.  Downward arrows represent the upper limits of radio luminosities for undetected cases in the NMA observations.  Left-pointing arrows represent the upper limits of X-ray luminosities.  NGC~3226, NGC~4168, NGC~4278, NGC~4374, NGC~4552, and NGC~5846, which are VLBI-detected LLAGNs of early-type galaxies, are plotted in both panels.  As a reference, Sgr~A* in a quiescent state and a flaring state are plotted in the left panel.  
\label{figure2}}
\end{figure*}

\subsection{Summary of results}
Among the 16~LLAGN samples, significant variability at 100~GHz was found in three LLAGNs (NGC~3718, NGC~4258, and NGC~4278).  Nuclear emissions at 100~GHz were detected in nine early-type galaxies (including three sources in the LLAGN sample) out of the 24~galaxies.  All detected sources have also been detected using VLBIs at lower frequencies, indicating that these centimeter emissions originate from compact nuclear components with high brightness temperatures ($\ga 10^8$~K, i.e., VLBI sensitivities).  The 100-GHz emissions found in several early-type galaxies cannot be responsible for the dust or extended nonthermal emissions.  They presumably originate from compact nuclear components with flat or inverted spectra.

\section{DISCUSSION}\label{section:discussion}

The implications of this study are that compact radio components reside in many nearby galaxies, and that their radio emissions show flat or inverted spectra up to at least 100~GHz and significant millimeter variability, which are radio properties similar to Sgr~A*.  We discuss the physical origin of the emissions detected at 100~GHz (and 150~GHz).  

\citet{Merloni_etal.2003} discovered ``a fundamental plane of black hole activity'' in three-dimensional space with all the ranges of black hole masses of $10^1$--$10^{10}M_{\sun}$ and X-ray luminosities of $10^{33}$--$10^{46}$~erg~s$^{-1}$ for the samples of X-ray binaries, our Galactic center, LLAGNs, low-ionization nuclear emission regions~(LINERs), Seyferts, radio galaxies, and radio-loud and radio-quiet quasars:  
\begin{equation}
L_\mathrm{5GHz} = 10^{7.33} L_\mathrm{X}^{0.6} M_\mathrm{BH}^{0.78}, \label{eq:fundamentalplane} 
\end{equation}
where $L_\mathrm{5GHz}$ is the nuclear radio luminosity at 5~GHz, $L_\mathrm{X}$ is the X-ray luminosity at 2--10~keV in erg~s$^{-1}$, and $M_\mathrm{BH}$ is the black hole mass in solar mass units \citep[see also][]{Falcke_etal.2004,Kording_etal.2006}.  One of the physical interpretations of this empirical relationship has been given by \citet{Heinz_Sunyaev2003} as a scale invariant disk--jet coupling of nonthermal jets powered by radiatively inefficient accretion flows onto black holes.  We compare the observed and empirically expected radio luminosities of our nearby LLAGN and early-type galaxy samples~(Figure~\ref{figure2}).  For the plots, the observed radio luminosities at 100~GHz are converted to those at 5~GHz, assuming a flat spectrum.  X-ray luminosities and black hole masses of our samples are listed in Tables~\ref{table1} and \ref{table2}.  Almost all LLAGNs appear to reside in accordance with the relationship.  The early-type galaxies are also in good agreement with the relation, indicating that the detected radio emissions are physically caused by active nuclei, i.e., nonthermal jets emanated from radiatively inefficient accretion flows, rather than any galactic components.  Considering the luminosities of these early-type galaxies, the active nuclei are presumably LLAGNs \citep[$L_\mathrm{X}<10^{42}$~ergs~s$^{-1}$; e.g.,][]{Terashima_Wilson2003} with starved SMBHs.  For Sgr~A* with $L_\mathrm{X}=2.2\times10^{33}$~erg~s$^{-1}$ in a quiescent state \citep{Baganoff_etal.2001} and $M_\mathrm{BH}=4\times10^6 M_\sun$ \citep{Gillessen_etal.2009}, a radio luminosity expected from the fundamental plane is $3.1 \times 10^{32}$~erg~s$^{-1}$, which is consistent with the observed radio luminosity of $2.5 \times 10^{32}$~erg~s$^{-1}$ \citep{Falcke_etal.1998} (cf.~$L_\mathrm{X}=1.0\times10^{35}$~erg~s$^{-1}$ in a flaring state \citep{Baganoff_etal.2001}).  Despite the difference by about 5~orders of magnitude in radio luminosity, Sgr~A* and our samples of nearby LLAGNs and early-type galaxies are on the same empirical law of the black hole activity.  The study for refining the fundamental plane by \citet{Kording_etal.2006} includes only objects in the low/hard state: LLAGNs, LINERs, FR-I radio galaxies \citep{Fanaroff_Riley1974}, BL~Lac objects, and X-ray binaries in the low/hard state.  The resultant parameters are ($\xi_\mathrm{R}$, $\xi_\mathrm{M}$, $b_\mathrm{X}$)=(1.41, $-0.87$, $-5.01$) corresponding to $L_\mathrm{5GHz} = 10^{3.55} L_\mathrm{X}^{0.71} M_\mathrm{BH}^{0.62}$, which provides expected radio luminosities $\sim 0.5--1$~dex lower than those by Equation~(\ref{eq:fundamentalplane}), i.e., moves points to leftward in the plots of Figure~\ref{figure2}.  The refined fundamental plane is also applicable or possibly better for our sample, if we consider many points as the upper limits of observed radio luminosities, possibly intrinsic inverted spectra between 5 and 100~GHz rather than flat spectra, and Sgr~A* in a flaring state.  The set of parameters is explained by synchrotron emissions at radio-through-X-rays in the jet-dominated state \citep[e.g.,][]{Falcke_Biermann1995}.  In either case, the picture that the detected millimeter radio emissions are caused by jets as the activity of SMBHs is suggested.     

There are two outliers, departed from the fundamental plane of Equation~(\ref{eq:fundamentalplane}), which may be highly variable radio sources at a quiescent and a flaring state.  The LLAGN NGC~3147 appears to have significantly low radio luminosity with respect to the fundamental plane (Figure~\ref{figure2}).  We speculate that the observed 100-GHz flux density was in a relatively quiescent state during our NMA observation.  NGC~3147 showed a slightly inverted spectrum at 1.7--43~GHz, which could extrapolate to $\sim$10--15~mJy at 100~GHz, three to five times larger than the NMA flux density of 3.4$\pm$1.5~mJy \citep[Figure~1 in][]{Doi_etal.2005b}.  On the other hand, the nucleus of the early-type galaxy NGC~5077 appears to have significantly high radio luminosity with respect to the fundamental plane (Figure~\ref{figure2}).  We speculate that the observed 100-GHz flux density was in a flaring state during our NMA observation, considering its flat or slightly inverted spectra in centimeters were extrapolated to 100~GHz.  Note that the employed X-ray luminosity is a ``pseudo-bolometric X-ray luminosity'' converted from the data of the {\it Einstein} IPC \citep{OSullivan_etal.2001}, which is an overestimated value of 2--10~keV X-ray luminosity, which results in an overpredicted radio luminosity.  

The ADAF model predicts inverted radio spectra on the basis of the synchrotron emissions from optically thick plasma with frequency-dependent radii, under the assumption of Maxwellian distribution of thermal electrons of $10^9$--$10^{10}$~K \citep{Narayan_etal.1995}.  Although this model explains a large portion of SED from millimeter wavelengths to hard X-rays, it has been difficult to explain the low-frequency radio luminosities \citep{Mahadevan1998}.  \citet{Wu_Cao2005} showed that the 5~GHz radio luminosities are higher than the maximal luminosities expected from the ADAF model for most sources in their LLAGN sample.  Although jets presumably dominate low-frequency radio emissions of LLAGNs, disk emissions of inverted spectra might predominate at millimeter and submillimeter wavelengths of very low luminosity sources \citep{Doi_etal.2005b}.  For a typical black hole with a mass of $10^8M_\sun$ and at a distance of 20~Mpc, the maximal luminosity and flux density calculated by the ADAF model \citep{Mahadevan1997} are $\sim 10^{38}$~erg~s$^{-1}$ and $\sim3$~mJy, respectively, at 100~GHz, which are near the detection limits of our observations.  All the detected radio powers could not be attributed to the thermal process in the ADAF.  

To produce a flat spectrum from the nonthermal synchrotron emission of electrons with a power-law energy distribution, self-absorbed multiple discrete components or continuous flows are required \citep[e.g.,][]{Blandford&Konigl1979}.  The emissions detected at 100~GHz are expected to mostly originate from spectral components with peaks at $\sim100$~GHz.  In the typical case that the emission of 10~mJy at the spectral peak originates at the distance of 20~Mpc, we consider a homogeneous spherical blob of synchrotron plasma under the equipartition between the energy densities of particles and magnetic fields.  The derived diameter of the blob is $\sim$70~AU, corresponding to the light-crossing time of $\sim$0.4~days, $\sim$35-times the Schwarzschild radius of the black hole with mass of $10^8 M_\sun$, or $\sim$4~$\mu$as.  Thus, intra-day variability is intrinsically explained for such emissions at 100~GHz.  The size of the emitting region is equivalent to the expected sizes of the radio emitting region of an accretion disk and that of jet production regions.  Thus, the millimeter emissions with flat or inverted spectra definitely originate from the activity in the vicinity of black holes.   

This physical consideration can be applied to the discussion of the possibility that the detected millimeter emissions originate from very young radio galaxies.  Their compact radio lobes are expected to show spectra peaked at high frequencies.  In the evolution of young radio-loud AGNs, the ratio of the equipartition component size to the overall linear size remains constant at 5--6, throughout the samples of faint and bright GHz-peaked spectrum~(GPS) and compact steep spectrum~(CSS) galaxies \citep{Snellen_etal.2000}.  If this self-similar evolution based on synchrotron self-absorption \citep{Fanti_etal.1990} is applied to extremely small radio sources, the expected overall size of the radio sources with a spectral peak of $\sim100$~GHz is $\sim500$~AU; this implies an age of two weeks by assuming a typical advancing speed of $\sim0.2c$ for the GPS and CSS galaxies \citep[e.g.,][]{Conway2002,Nagai_etal.2006}.  The lifetime of the activity predicts a very small chance for such extremely young radio sources to be found in the limited number of samples.  NGC~4278, one of our LLAGN samples, is a low-power compact source \citep[LPC;][]{Giroletti_etal.2005b}, and its measured apparent velocity of jets indicates the age of 8.3--65.8~years \citep{Giroletti_etal.2005a}.  Although NGC~4278 might be a young radio galaxy, it is old enough to show its steep spectrum.

\section{Summary}\label{section:summary}
We conducted millimeter continuum observations of the central regions of nearby LLAGNs and early-type galaxies using the NMA.  The results and their implications are described below.   
\begin{itemize}
\item Significant variability at 100~GHz was found in 3 out of 16 ($\sim 19$\%) LLAGNs (NGC~3718, NGC~4258, and NGC~4278) by reanalyzing the data of a previous study \citep{Doi_etal.2005b}.  Significant variability was shown in NGC~4278 in only two days.    
\item The flux measurements by the new NMA quasi-simultaneous observations at 100 and 150~GHz were presented for five LLAGNs (NGC~3031~(M~81), NGC~4278, NGC~4374~(M84), NGC~4579~(M58), and NGC~6500).      
\item From the new NMA observations, nuclear emissions at 100~GHz were detected in nine~early-type galaxies (including three sources in the LLAGN sample) out of the 24~galaxies ($\sim 38$\%).  
\item The detected 100~GHz emissions in many LLAGNs and several early-type galaxies cannot be responsible for the dust or extended nonthermal emissions; these emissions are presumably from compact nuclear components, similar to Sgr~A*, with flat or inverted spectra at centimeter-to-millimeter wavelengths.   
\item The observed radio luminosities are consistent with the fundamental plane of black hole activity \citep{Merloni_etal.2003}; thus, implying compact nuclear jets emanating from radiatively inefficient accretion flows as the physical origin.  
\end{itemize}

\acknowledgments
We acknowledge the support of the NRO staff in operating the telescopes and for their continuous efforts to improve the performance of the instruments.  The NRO is a branch of the National Astronomical Observatory of Japan~(NAOJ), which belongs to National Institutes of Natural Sciences (NINS).  We also acknowledge S.~Okumura for helpful advices in observations with the NMA at the NRO.  We have used NASA's Astrophysics Data System Abstract Service, the NASA/IPAC Extragalactic Database (NED), which is operated by the Jet Propulsion Laboratory.   This work was partially supported by a Grant-in-Aid for Young Scientists~(B; 18740107, A.~D.) and a Grant-in-Aid for Scientific Research~(C; 21540250, A.~D.) from the Japanese Ministry of Education, Culture, Sports, Science, and Technology~(MEXT).  This work was supported in part by the Center for the Promotion of Integrated Sciences~(CPIS) of Sokendai.


\setcounter{figure}{0}
\clearpage
\begin{landscape}
\begin{figure*}
\epsscale{1.15}
\plotone{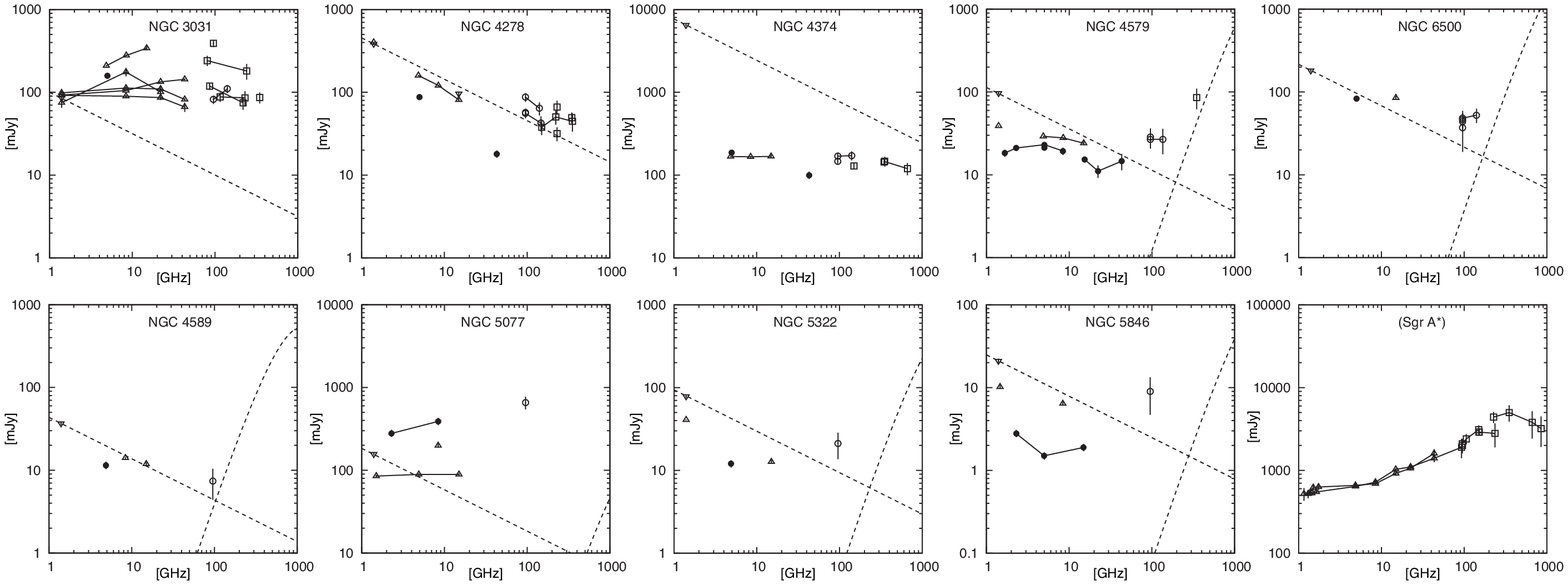}
\figcaption{
Radio spectra of LLAGNs observed at both $\sim100$~GHz~($\lambda3$mm) and $\sim150$~GHz~($\lambda2$mm) and those of early-type galaxies detected at $\sim100$~GHz.  As a reference, the radio spectra of Sgr~A* is also presented.  {\bf Symbols connected with solid lines} represent quasi-simultaneous observations.  {\bf Open circles} represent our NMA measurements at $\sim100$~GHz and $\sim150$~GHz.  {\bf Open squares} show millimeter single-dish and millimeter interferometric measurements (NGC~3031: \citealt{Doi_etal.2005b,Schodel_etal.2007}, NGC~4278: \citealt{Anton_etal.2004,Crocker_etal.2011}, NGC~4374: \citealt{Leeuw_etal.2004}, NGC~4579: \citealt{Doi_etal.2005b}, Sgr~A*: \citealt{Marrone_etal.2008}).  {\bf Downward triangles at 1.4~GHz} represent VLA-D measurements \citep[the NRAO VLA Sky Survey: NVSS][]{Condon_etal.1998} or single dish \citep[NGC~4374:][]{White&Becker1992}.  {\bf Upward triangles} at 1.4~GHz are VLA-B measurements \citep[Faint Images of the Radio Sky at Twenty centimeters: FIRST,][]{Becker_etal.1995}.  {\bf Connected upward triangles} represent simultaneous VLA-A measurements at 5, 8.4, and 15~GHz \citep{Nagar_etal.2001} except for NGC~5077 \citep{Wrobel&Heeschen1984} and VLA-B or -C at 1.4--22~GHz for NGC~3031 \citep{Markoff_etal.2008}.  {\bf The other upward triangles} are VLA-A measurements for NGC~6500 \citep{Nagar_etal.2005}, NGC~4589 \citep{Jackson_etal.2007, Nagar_etal.2005}, NGC~5077 \citep{Healey_etal.2007}, NGC~5322 \citep{Nagar_etal.2005}, NGC~5846 \citep{Dunn_etal.2010,Filho_etal.2004}, and Sgr~A* \citep{An_etal.2005,Falcke_etal.1998}.  {\bf Filled circles} are VLBI measurements (NGC~3031: \citealt{Bartel&Bietenholtz2000}, NGC~4278: \citealt{Falcke_etal.2000,Ly_etal.2004}, NGC~4374: \citealt{Nagar_etal.2005,Ly_etal.2004}, NGC~4579: \citealt{Falcke_etal.2000,Anderson_etal.2004,Ulvestad&Ho2001}, NGC~6500: \citealt{Falcke_etal.2000}, NGC~4589: \citealt{Nagar_etal.2005}, NGC~5077: \citealt{Petrov_etal.2006}, NGC~5322: \citealt{Nagar_etal.2005}, and NGC~5846: \citealt{Filho_etal.2004}).  {\bf Dashed lines of the dependence of $\nu^{-0.5}$} (where $\nu$ is frequency) represent upper limits on the possible contribution from extended nonthermal emission, constrained by 1.4-GHz data obtained with beam sizes much larger than that of NMA.  {\bf Dashed lines of highly inverted spectra} represent upper limits of the possible contribution from dust emission, determined by the following data obtained with beam sizes much larger than that of NMA: one or two-temperature model-fitting with $\beta=1$ to {\it Spitzer/ISO/IRAS} measurements at 24--200~$\mu$m (NGC~6500: \citealt{Stickel_etal.2004,1990IRASF.C......0M}, NGC~4589: \citealt{Temi_etal.2004,Temi_etal.2007,Kaneda_etal.2007}, NGC~5077: \citealt{Temi_etal.2007}, NGC~5322: \citealt{Temi_etal.2004,Temi_etal.2007}, NGC~5846: \citealt{Temi_etal.2007}).  For NGC~4579, the measurements with a 17\arcsec aperture based on our analyses of {\it Spitzer} archival data.           
\label{figure1}}
\end{figure*}
\clearpage
\end{landscape}


\begin{thebibliography}{}
\bibitem[An et al.(2005)]{An_etal.2005} An, T., Goss, W.~M., Zhao, J.-H., Hong, X.~Y., Roy, S., Rao, A.~P., \& Shen, Z.-Q.\ 2005, \apjl, 634, L49

\bibitem[Anderson et al.(2004)]{Anderson_etal.2004} Anderson, J.~M., Ulvestad, J.~S., \& Ho, L.~C.\ 2004, \apj, 603, 42 

\bibitem[Ant{\'o}n et al.(2004)]{Anton_etal.2004} Ant{\'o}n, S., Browne, I.~W.~A., March{\~a}, M.~J.~M., Bondi, M., \& Polatidis, A.\ 2004, \mnras, 352, 673 


\bibitem[Baganoff et al.(2001)]{Baganoff_etal.2001} Baganoff, F.~K., et al.\ 2001, \nat, 413, 45 

\bibitem[Bartel \& Bietenholz(2000)]{Bartel&Bietenholtz2000} Bartel, N., \& Bietenholz, M.~F.\ 2000, Astrophysical Phenomena Revealed by Space VLBI, 17 

\bibitem[Becker et al.(1995)]{Becker_etal.1995} Becker, R.~H., White, R.~L., \& Helfand, D.~J.\ 1995, \apj, 450, 559 

\bibitem[Bender et al.(1994)]{Bender_etal.1994} Bender, R., Saglia, R.~P., \& Gerhard, O.~E.\ 1994, \mnras, 269, 785 

\bibitem[Bernardi et al.(2002)]{Bernardi_etal.2002} Bernardi, M., Alonso, M.~V., da Costa, L.~N., Willmer, C.~N.~A., Wegner, G., Pellegrini, P.~S., Rit{\'e}, C., \& Maia, M.~A.~G.\ 2002, \aj, 123, 2990 

\bibitem[Bettoni \& Buson(1987)]{Bettoni_Buson1987} Bettoni, D., \& Buson, L.~M.\ 1987, \aaps, 67, 341 

\bibitem[Blandford \& Konigl(1979)]{Blandford&Konigl1979} Blandford, R.~D., \& Konigl, A.\ 1979, \apj, 232, 34 

\bibitem[Bower et al.(1998)]{Bower_etal.1998} Bower, G.~A., et al.\ 1998, \apjl, 492, L111 


\bibitem[Brown et al.(2011)]{Brown_etal.2011} Brown, M.~J.~I., Jannuzi, B.~T., Floyd, D.~J.~E., \& Mould, J.~R.\ 2011, \apjl, 731, L41 

\bibitem[Condon et al.(1998)]{Condon_etal.1998} Condon, J.~J., Cotton, W.~D., Greisen, E.~W., Yin, Q.~F., Perley, R.~A., Taylor, G.~B., \& Broderick, J.~J.\ 1998, \aj, 115, 1693 

\bibitem[Conway(2002)]{Conway2002} Conway, J.~E.\ 2002, NewAR, 46, 263 

\bibitem[Crocker et al.(2008)]{Crocker_etal.2008} Crocker, A.~F., Bureau, M., Young, L.~M., \& Combes, F.\ 2008, \mnras, 386, 1811

\bibitem[Crocker et al.(2011)]{Crocker_etal.2011} Crocker, A.~F., Bureau, M., Young, L.~M., \& Combes, F.\ 2011, \mnras, 410, 1197




\bibitem[Desroches \& Ho(2009)]{Desroches_Ho2009} Desroches, L.-B., \& Ho, L.~C.\ 2009, \apj, 690, 267 

\bibitem[Devereux et al.(2003)]{Devereux_etal.2003} Devereux, N., Ford, H., Tsvetanov, Z., \& Jacoby, G.\ 2003, \aj, 125, 1226 

\bibitem[Doi et al.(2005a)]{Doi_etal.2005a} Doi, A., Kameno, S., 
\& Inoue, M.\ 2005a, \mnras, 360, 119 

\bibitem[Doi et al.(2005b)]{Doi_etal.2005b} Doi, A., Kameno, S., Kohno, K., Nakanishi, K., \& Inoue, M.\ 2005b, \mnras, 363, 692

\bibitem[Doi et al.(2011)]{Doi_etal.2011} Doi, A., Kohno, K., Nakanishi, K., Kameno, S., Inoue, M., Hada, K., \& Sorai, K.\ 2011, arXiv:1106.2930 

\bibitem[Dunn et al.(2010)]{Dunn_etal.2010} Dunn, R.~J.~H., Allen, S.~W., Taylor, G.~B., Shurkin, K.~F., Gentile, G., Fabian, A.~C., \& Reynolds, C.~S.\ 2010, \mnras, 404, 180 

\bibitem[Eracleous et al.(2002)]{Eracleous_etal.2002} Eracleous, M., Shields, J.~C., Chartas, G., \& Moran, E.~C.\ 2002, \apj, 565, 108 

\bibitem[Eracleous et al.(2010)]{Eracleous_etal.2010} Eracleous, M., Hwang, J.~A., \& Flohic, H.~M.~L.~G.\ 2010, \apjs, 187, 135 

\bibitem[Falcke \& Biermann(1995)]{Falcke_Biermann1995} Falcke, H., \& Biermann, P.~L.\ 1995, \aap, 293, 665 



\bibitem[Falcke et al.(1998)]{Falcke_etal.1998} Falcke, H., Goss, W.~M., Matsuo, H., Teuben, P., Zhao, J.-H., \& Zylka, R.\ 1998, \apj, 499, 731 

\bibitem[Falcke et al.(2000)]{Falcke_etal.2000} Falcke, H., Nagar, N.~M., Wilson, A.~S., \& Ulvestad, J.~S.\ 2000, \apj, 542, 197 

\bibitem[Falcke et al.(2004)]{Falcke_etal.2004} Falcke, H., K{\"o}rding, E., \& Markoff, S.\ 2004, \aap, 414, 895

\bibitem[Fanaroff \& Riley(1974)]{Fanaroff_Riley1974} Fanaroff, B.~L., \& Riley, J.~M.\ 1974, \mnras, 167, 31P 

\bibitem[Fanti et al.(1990)]{Fanti_etal.1990} Fanti, R., Fanti, C., Schilizzi, R.~T., Spencer, R.~E., Nan Rendong, Parma, P., van Breugel, W.~J.~M., \& Venturi, T.\ 1990, \aap, 231, 333

\bibitem[Filho et al.(2004)]{Filho_etal.2004} Filho, M.~E., Fraternali, F., Markoff, S., Nagar, N.~M., Barthel, P.~D., Ho, L.~C., \& Yuan, F.\ 2004, \aap, 418, 429 

\bibitem[Freedman et al.(1994)]{Freedman_etal.1994} Freedman, W.~L., et al.\ 1994, \apj, 427, 628 

\bibitem[Gebhardt et al.(2003)]{Gebhardt_etal.2003} Gebhardt, K., et al.\ 2003, \apj, 583, 92 

\bibitem[Ghosh et al.(2007)]{Ghosh_etal.2007} Ghosh, H., Pogge, R.~W., 
Mathur, S., Martini, P., \& Shields, J.~C.\ 2007, \apj, 656, 105 

\bibitem[Gillessen et al.(2009)]{Gillessen_etal.2009} Gillessen, S., Eisenhauer, F., Trippe, S., Alexander, T., Genzel, R., Martins, F., \& Ott, T.\ 2009, \apj, 692, 1075 

\bibitem[Giroletti et al.(2005a)]{Giroletti_etal.2005a} Giroletti, M., 
Taylor, G.~B., \& Giovannini, G.\ 2005a, \apj, 622, 178 

\bibitem[Giroletti et al.(2005b)]{Giroletti_etal.2005b} Giroletti, M., Giovannini, G., \& Taylor, G.~B.\ 2005b, \aap, 441, 89 

\bibitem[Gonz{\'a}lez(1993)]{Gonzalez1993} Gonz{\'a}lez, J.~J.\ 1993, Ph.D.~Thesis,  

\bibitem[Gonz{\'a}lez-Mart{\'{\i}}n et al.(2006)]{Gonzalez-Martin_etal.2006} Gonz{\'a}lez-Mart{\'{\i}}n, O., Masegosa, J., M{\'a}rquez, I., Guerrero, M.~A., \& Dultzin-Hacyan, D.\ 2006, \aap, 460, 45 


\bibitem[Goudfrooij et al.(1994)]{Goudfrooij_etal.1994} Goudfrooij, P., Hansen, L., Jorgensen, H.~E., \& Norgaard-Nielsen, H.~U.\ 1994, \aaps, 105, 341


\bibitem[G{\"u}ltekin et al.(2009)]{Gultekin_etal.2009} G{\"u}ltekin, K., Cackett, E.~M., Miller, J.~M., Di Matteo, T., Markoff, S., \& Richstone, D.~O.\ 2009, \apj, 706, 404 

\bibitem[Halderson et al.(2001)]{Halderson_etal.2001} Halderson, E.~L., Moran, E.~C., Filippenko, A.~V., \& Ho, L.~C.\ 2001, \aj, 122, 637 

\bibitem[Healey et al.(2007)]{Healey_etal.2007} Healey, S.~E., Romani, R.~W., Taylor, G.~B., Sadler, E.~M., Ricci, R., Murphy, T., Ulvestad, J.~S., \& Winn, J.~N.\ 2007, \apjs, 171, 

\bibitem[Heckman et al.(1980)]{Heckman_etal.1980} Heckman, T.~M., Crane, P.~C., \& Balick, B.\ 1980, \aaps, 40, 295 

\bibitem[Heinz \& Sunyaev(2003)]{Heinz_Sunyaev2003} Heinz, S., \& Sunyaev, R.~A.\ 2003, \mnras, 343, L59 

\bibitem[Herrnstein et al.(1997)]{Herrnstein_etal.1997} Herrnstein, J.~R., Moran, J.~M., Greenhill, L.~J., Diamond, P.~J., Miyoshi, M., Nakai, N., \& Inoue, M.\ 1997, \apjl, 475, L17 

\bibitem[Herrnstein et al.(1999)]{Herrnstein_etal.1999} Herrnstein, J.~R., et al.\ 1999, \nat, 400, 539 

\bibitem[Ho et al.(1997a)]{Ho_etal.1997a} Ho, L.~C., Filippenko, A.~V., \& Sargent, W.~L.~W.\ 1997a, \apj, 487, 568 

\bibitem[Ho et al.(1997b)]{Ho_etal.1997b} Ho, L.~C., Filippenko, 
A.~V., \& Sargent, W.~L.~W.\ 1997b, \apjs, 112, 315 

\bibitem[Ho et al.(1999)]{Ho_etal.1999} Ho, L.~C., van Dyk, S.~D., 
Pooley, G.~G., Sramek, R.~A., \& Weiler, K.~W.\ 1999, \aj, 118, 843 


\bibitem[Ho et al.(2001)]{Ho_etal.2001} Ho, L.~C., et al.\ 2001, \apjl, 549, L51 

\bibitem[Ho(2007)]{Ho2007} Ho, L.~C.\ 2007, \apj, 668, 94 

\bibitem[Huchra \& Burg(1992)]{Huchra_Burg1992} Huchra, J., \& Burg, R.\ 1992, \apj, 393, 90 

\bibitem[Hyman et al.(2001)]{Hyman_etal.2001} Hyman, S.~D., Calle, D., Weiler, K.~W., Lacey, C.~K., Van Dyk,  S.~D., \& Sramek, R.\ 2001, \apj, 551, 702 

\bibitem[Iyomoto et al.(1998)]{Iyomoto_etal.1998} Iyomoto, N., Makishima, K., Matsushita, K., Fukazawa, Y., Tashiro, M., \& Ohashi, T.\ 1998, \apj, 503, 168 

\bibitem[Jackson et al.(2007)]{Jackson_etal.2007} Jackson, N., Battye, 
R.~A., Browne, I.~W.~A., Joshi, S., Muxlow, T.~W.~B., \& Wilkinson, P.~N.\ 2007, \mnras, 376, 371

\bibitem[Kaneda et al.(2007)]{Kaneda_etal.2007} Kaneda, H., Onaka, T., Kitayama, T., Okada, Y., \& Sakon, I.\ 2007, \pasj, 59, 107

\bibitem[Kollatschny \& Fricke(1989)]{Kollatschny_Fricke1989} Kollatschny, W., \& Fricke, K.~J.\ 1989, \aap, 219, 34 

\bibitem[Komossa et al.(1999)]{Komossa_etal.1999} Komossa, S., B{\"o}hringer, H., \& Huchra, J.~P.\ 1999, \aap, 349, 88 


\bibitem[K{\"o}rding et al.(2006)]{Kording_etal.2006} K{\"o}rding, E., Falcke, H., \& Corbel, S.\ 2006, \aap, 456, 439 

\bibitem[Kormendy \& Gebhardt(2001)]{Kormendy_Gebhardt2001} Kormendy, J., \& Gebhardt, K.\ 2001, 20th Texas Symposium on relativistic astrophysics, 586, 363 

\bibitem[Krips et al.(2005)]{Krips_etal.2005} Krips, M., et al.\ 2005, \aap, 442, 479 

\bibitem[Kuntschner et al.(2001)]{Kuntschner_etal.2001} Kuntschner, H., Lucey, J.~R., Smith, R.~J., Hudson, M.~J., \& Davies, R.~L.\ 2001, \mnras, 323, 615 

\bibitem[Leeuw et al.(2004)]{Leeuw_etal.2004} Leeuw, L.~L., Sansom, A.~E., Robson, E.~I., Haas, M., \& Kuno, N.\ 2004, \apj, 612, 83

\bibitem[Liu \& Bregman(2005)]{Liu_Bregman2005} Liu, J.-F., \& Bregman, J.~N.\ 2005, \apjs, 157, 59 

\bibitem[Ly et al.(2004)]{Ly_etal.2004} Ly, C., Walker, R.~C., \& Wrobel, J.~M.\ 2004, \aj, 127, 119

\bibitem[Mahadevan(1997)]{Mahadevan1997} Mahadevan, R.\ 1997, \apj, 477, 585 

\bibitem[Mahadevan(1998)]{Mahadevan1998} Mahadevan, R.\ 1998, \nat, 394, 651 

\bibitem[Manmoto et al.(1997)]{Manmoto_etal.1997} Manmoto, T., Mineshige, S., \& Kusunose, M.\ 1997, \apj, 489, 791 

\bibitem[Marconi et al.(2004)]{Marconi_etal.2004} Marconi, A., Risaliti, G., Gilli, R., Hunt, L.~K., Maiolino, R., \& Salvati, M.\ 2004, \mnras, 351, 169

\bibitem[Markoff et al.(2008)]{Markoff_etal.2008} Markoff, S., et al.\ 2008, \apj, 681, 905 

\bibitem[Marrone et al.(2008)]{Marrone_etal.2008} Marrone, D.~P., et al.\ 2008, \apj, 682, 373 


\bibitem[McElroy(1995)]{McElroy1995} McElroy, D.~B.\ 1995, \apjs, 100, 105 

\bibitem[Merloni et al.(2003)]{Merloni_etal.2003} Merloni, A., Heinz, S., \& di Matteo, T.\ 2003, \mnras, 345, 1057

\bibitem[Miyazaki et al.(2004)]{Miyazaki_etal.2004} Miyazaki, A., Tsutsumi, T., \& Tsuboi, M.\ 2004, \apjl, 611, L97 

\bibitem[Moshir \& et al.(1990)]{1990IRASF.C......0M} Moshir, M., \& et al.\ 1990, IRAS Faint Source Catalogue, version 2.0 (1990), 

\bibitem[Nagai et al.(2006)]{Nagai_etal.2006} Nagai, H., Inoue, M., Asada, K., Kameno, S., \& Doi, A.\ 2006, \apj, 648, 148 

\bibitem[Nagar et al.(2001)]{Nagar_etal.2001} Nagar, N.~M., Wilson, 
A.~S., \& Falcke, H.\ 2001, \apjl, 559, L87

\bibitem[Nagar et al.(2005)]{Nagar_etal.2005} Nagar, N.~M., Falcke, H., \& Wilson, A.~S.\ 2005, \aap, 435, 521 

\bibitem[Narayan et al.(1995)]{Narayan_etal.1995} Narayan, R., Yi, I., \& Mahadevan, R.\ 1995, \nat, 374, 623 


\bibitem[Okumura et al.(2000)]{Okumura_etal.2000} Okumura, S.~K., et al.\ 2000, \pasj, 52, 393 



\bibitem[O'Sullivan et al.(2001)]{OSullivan_etal.2001} O'Sullivan, E., Forbes, D.~A., \& Ponman, T.~J.\ 2001, \mnras, 328, 461 

\bibitem[Panessa et al.(2006)]{Panessa_etal.2006} Panessa, F., Bassani, L., Cappi, M., Dadina, M., Barcons, X., Carrera, F.~J., Ho, L.~C., \& Iwasawa, K.\ 2006, \aap, 455, 173 

\bibitem[Pellegrini(2005)]{Pellegrini2005} Pellegrini, S.\ 2005, \apj, 
624, 155 

\bibitem[Pellegrini(2010)]{Pellegrini2010} Pellegrini, S.\ 2010, \apj, 717, 640 

\bibitem[Rest et al.(2001)]{Rest_etal.2001} Rest, A., van den Bosch, F.~C., Jaffe, W., Tran, H., Tsvetanov, Z., Ford, H.~C., Davies, J., \& Schafer, J.\ 2001, \aj, 121, 2431 

\bibitem[Petrov et al.(2006)]{Petrov_etal.2006} Petrov, L., Kovalev, Y.~Y., Fomalont, E.~B., \& Gordon, D.\ 2006, \aj, 131, 1872 

\bibitem[Reuter \& Lesch(1996)]{Reuter&Lesch1996} Reuter, H.-P., \& Lesch, H.\ 1996, \aap, 310, L5 


\bibitem[Reynolds et al.(2009)]{Reynolds_etal.2009} Reynolds, C.~S., Nowak, M.~A., Markoff, S., Tueller, J., Wilms, J., \& Young, A.~J.\ 2009, \apj, 691, 1159 

\bibitem[Roberts et al.(1991)]{Roberts_etal.1991} Roberts, M.~S., Hogg, D.~E., Bregman, J.~N., Forman, W.~R., \& Jones, C.\ 1991, \apjs, 75, 751 

\bibitem[Sadler et al.(1989)]{Sadler_etal.1989} Sadler, E.~M., Jenkins, C.~R., \& Kotanyi, C.~G.\ 1989, \mnras, 240, 591 

\bibitem[Sakamoto et al.(2001)]{Sakamoto_etal.2001} Sakamoto, K., Fukuda, H., Wada, K., \& Habe, A.\ 2001, \aj, 122, 1319 

\bibitem[Sarzi et al.(2001)]{Sarzi_etal.2001} Sarzi, M., Rix, H.-W., Shields, J.~C., Rudnick, G., Ho, L.~C., McIntosh, D.~H., Filippenko, A.~V., \& Sargent, W.~L.~W.\ 2001, \apj, 550, 65 



\bibitem[Satyapal et al.(2005)]{Satyapal_etal.2005} Satyapal, S., Dudik, 
R.~P., O'Halloran, B., \& Gliozzi, M.\ 2005, \apj, 633, 86 

\bibitem[Sch{\"o}del et al.(2007)]{Schodel_etal.2007} Sch{\"o}del, R., Krips, M., Markoff, S., Neri, R., \& Eckart, A.\ 2007, \aap, 463, 551 

\bibitem[Slee et al.(1994)]{Slee_etal.1994} Slee, O.~B., Sadler, E.~M., Reynolds, J.~E., \& Ekers, R.~D.\ 1994, \mnras, 269, 928

\bibitem[Snellen et al.(2000)]{Snellen_etal.2000} Snellen, I.~A.~G., Schilizzi, R.~T., Miley, G.~K., de Bruyn, A.~G., Bremer, M.~N., R\"{o}ttgering, H.~J.~A.\ 2000, \mnras, 319, 445 



\bibitem[Solanes et al.(2002)]{Solanes_etal.2002} Solanes, J.~M., Sanchis, T., Salvador-Sol{\'e}, E., Giovanelli, R., \& Haynes, M.~P.\ 2002, \aj, 124, 2440 

\bibitem[Soria et al.(2006)]{Soria_etal.2006} Soria, R., Fabbiano, G., Graham, A.~W., Baldi, A., Elvis, M., Jerjen, H., Pellegrini, S., \& Siemiginowska, A.\ 2006, \apj, 640, 126 

\bibitem[Stickel et al.(2004)]{Stickel_etal.2004} Stickel, M., Lemke, D., Klaas, U., Krause, O., \& Egner, S.\ 2004, \aap, 422, 39

\bibitem[Temi et al.(2004)]{Temi_etal.2004} Temi, P., Brighenti, F., Mathews, W.~G., \& Bregman, J.~D.\ 2004, \apjs, 151, 237 

\bibitem[Temi et al.(2007)]{Temi_etal.2007} Temi, P., Brighenti, F., 
\& Mathews, W.~G.\ 2007, \apj, 660, 1215 

\bibitem[Terashima \& Wilson(2003)]{Terashima_Wilson2003} Terashima, Y., \& Wilson, A.~S.\ 2003, \apj, 583, 145 

\bibitem[Tonry et al.(2001)]{Tonry_etal.2001} Tonry, J.~L., Dressler, A., Blakeslee, J.~P., Ajhar, E.~A., Fletcher, A.~B., Luppino, G.~A., Metzger, M.~R., \& Moore, C.~B.\ 2001, \apj, 546, 681 

\bibitem[Tran et al.(2001)]{Tran_etal.2001} Tran, H.~D., Tsvetanov, Z., Ford, H.~C., Davies, J., Jaffe, W., van den Bosch, F.~C., \& Rest, A.\ 2001, \aj, 121, 2928 

\bibitem[Tremaine et al.(2002)]{Tremaine+2002} Tremaine, S., et al.\ 2002, \apj, 574, 740 

\bibitem[Tsutsumi et al.(1997)]{Tsutsumi_etal.1997} Tsutsumi, T., Morita, 
K.-I., \& Umeyama, S.\ 1997, Astronomical Data Analysis Software and Systems VI, 125, 50

\bibitem[Tully(1988)]{Tully1988} Tully, R.~B.\ 1988, Cambridge and New York, Cambridge University Press, 1988, 221 p., 

\bibitem[Ulvestad \& Ho(2001)]{Ulvestad&Ho2001} Ulvestad, J.~S., \& Ho, L.~C.\ 2001, \apjl, 562, L133

\bibitem[van der Kruit et al.(1972)]{vanderKruit_etal.1972} van der Kruit, P.~C., Oort, J.~H., \& Mathewson, D.~S.\ 1972, \aap, 21, 169 


\bibitem[Wegner et al.(2003)]{Wegner_etal.2003} Wegner, G., et al.\ 2003, \aj, 126, 2268 

\bibitem[White \& Becker(1992)]{White&Becker1992} White, R.~L., \& Becker, R.~H.\ 1992, \apjs, 79, 331

\bibitem[Wrobel \& Heeschen(1984)]{Wrobel&Heeschen1984} Wrobel, J.~M., \& Heeschen, D.~S.\ 1984, \apj, 287, 41 

\bibitem[Wu \& Cao(2005)]{Wu_Cao2005} Wu, Q., \& Cao, X.\ 2005, \apj, 621, 130 



\bibitem[Yuan et al.(2003)]{Yuan_etal.2003} Yuan, F., Quataert, E., \& Narayan, R.\ 2003, \apj, 598, 301 

\bibitem[Yusef-Zadeh et al.(2011)]{Yusef-Zadeh_etal.2011} Yusef-Zadeh, F., Wardle, M., Miller-Jones, J.~C.~A., Roberts, D.~A., Grosso, N., \& Porquet, D.\ 2011, \apj, 729, 44 

\end{thebibliography}
\end{document}